\documentclass[12pt]{article}
\usepackage[utf8]{inputenc}
\usepackage[margin=1.25in]{geometry}
\usepackage{graphicx,subcaption,lipsum}
\usepackage{comment}
\usepackage{float}
\usepackage{graphicx}
\usepackage{amsmath}
\usepackage{authblk}
\usepackage{eufrak}
\usepackage{mathtools}
\usepackage{tabularx}
\usepackage{booktabs}
\usepackage{natbib}
\usepackage{amsfonts}
\usepackage{hyperref}
\usepackage{amsmath}
\usepackage{appendix}

\DeclareMathOperator*{\argmin}{arg\,min}
\usepackage{array}
\setcitestyle{notesep={: }}
\setcitestyle{aysep={}} 
\newcommand{\proglang}[1]{{\normalfont\fontseries{b}\selectfont #1}}
\let\proglang=\textsf

\title{On the Improvement of Predictive Modeling Using Bayesian Stacking and Posterior Predictive Checking  \\[1ex] \large 
An Example of Modeling Gender Inequality in Reading Using PISA 2018}

\author[1]{Mariana Nold}
\author[2]{Florian Meinfelder}
\author[3]{David Kaplan}
\affil[1]{University of Jena}
\affil[2]{University of Bamberg}
\affil[3]{University of Wisconsin - Madison}
\begin{document}

\maketitle

\newpage
\begin{abstract}

Model uncertainty is pervasive in real world analysis situations and is an often-neglected issue in applied statistics.
However, standard approaches to the research process do not address the inherent uncertainty in model building and, thus, can lead to overconfident and misleading analysis interpretations. One strategy to incorporate more flexible models is to base inferences on predictive modeling. This approach provides an alternative to existing explanatory models, as inference is focused on the posterior predictive distribution of the response variable. Predictive modeling can advance explanatory ambitions in the social sciences and in addition enrich the understanding of social phenomena under investigation. Bayesian stacking is a methodological approach rooted in Bayesian predictive modeling. In this paper, we outline the method of Bayesian stacking but add to it the approach of posterior predictive checking (PPC) as a means of assessing the predictive quality of those elements of the stacking ensemble that are important to the research question. Thus, we introduce a viable workflow for incorporating PPC into predictive modeling using Bayesian stacking without presuming the existence of a true model. We apply these tools to the PISA 2018 data to investigate potential inequalities in reading competency with respect to gender and socio-economic background.  Our empirical example serves as rough guideline for practitioners who want to implement the concepts of predictive modeling and model uncertainty in their work to similar research questions.

\end{abstract}

 \section{Introduction}

 \par The relationship of a substantive research question and the statistical model used to address the question using survey data is an intricate process with many constraints. Data quality problems of surveys aside, such as measurement errors, or unit/item nonresponse, we also face constraints on a measurement level, i.e. the survey that serves as a data base might not contain precisely the variables our model ideally requires, or they are not quite measured in a way that reliably and validly assesses the construct of interest.
Research questions are also reduced to a mathematical form, where some response variable $Y$ is represented as a function of some explanatory variables $X$. Since this relationship is never perfect, a stochastic component  is additionally required to capture omitted variables and simple random noise. The field of statistics has provided the scientific community with a toolbox to estimate and interpret these models, but statistical theory implicitly assumes that the selected model, upon which all subsequent inferences are built, is "correct", and the only source of uncertainty is the stochastic nature of the stochastic component of the model and the fact that our data are a random sample from a population. However, this selected or final model is typically the result of an ad hoc trial-and-error process, where several models were investigated and discarded. This process is often subjective and not completely transparent, but we think it is fair to say that the choice of the final model is based on a mixture of pre-conceived ideas, i.e. the relationship between variables we are interested in (our "scientific hypothesis"), variables that serve as "control variables" (based on some exploratory analysis and some variable selection scheme, constrained by the availability of these control variables in the data set), and overall explanatory power or "significance" of the parameter estimates we are interested in. 
\par As part of the research process, we might have estimated dozens of models with different subsets of $X$, different transformations and interactions, and maybe even tried out different model classes. Sometimes, there is no clear-cut "winner", and the final model might have barely come out on top, based on an ad hoc decision, impacted by overall goodness-of-fit, or hard-to-define "plausibility". Clearly, there often is a considerable amount of uncertainty involved in the modeling process, whether or not the final model is the "right" or "correct" model, but it is not common practice to try to incorporate this uncertainty into inferential statistical analysis. \citet{breiman_little_1992} described this procedure as a "quiet scandal", and several publications \citep[see e.g.][]{chatfield_model_1995} have cited this statement over the years.  Regardless, the practice of ignoring this source of uncertainty and to follow the described ad hoc process is still widespread \citep*{vacheron_quiet_2021}. 
This reveals a gap between the progress of statistical theory and research practice. 
\par The Bayesian paradigm facilitates conditioning inferences on some model, and a first intuitive approach to incorporate model uncertainty was \emph{Bayesian model averaging} (BMA) \citep*{madiganraftery94, raftery_bayesian_1995, raftery97}. In general, techniques for incorporating model uncertainty have been discussed in the statistical community, but there is a shortage of actual real-world applications that incorporate these approaches in sociology \citep{LynchBartlett}. Our aim is to reduce this gap in the literature, because we believe that addressing model uncertainty is necessary, though not sufficient, to establish the credibility of statistical methods in empirical practice.

\subsection{Sources of Uncertainty in Statistical Modeling}\label{subsec:uncert}

\par One strategy to incorporate more flexible models in general is to base inferences on predictive modeling rather than direct interpretation of parameter estimates \citep{GalitShmueli.2010}. This approach provides an alternative to existing explanatory models, as inference is focused on the posterior predictive distribution of the response variable. 

The method of \emph{Bayesian stacking}  \citep*{Wolpert1992,LBreiman.1996,YulingYao.2018} is one methodological approach rooted in Bayesian predictive modeling, where posterior predictions from many complex candidate models are combined. It can be considered an alternative to BMA as a mean to account for model uncertainty that requires less restrictive assumptions. The term "model uncertainty" refers to all decisions within the specification of the statistical models. Additionally, we distinguish between predictive and inferential uncertainty \citep*[see][115]{Gelman.2021}: Inferential uncertainty describes uncertainty due to the estimation of the parameters, whereas predictive uncertainty refers to the stochastic uncertainty involved in the prediction process. If we base our workflow on predictive modeling via Bayesian stacking we can account for these aforementioned sources of uncertainty in a natural and coherent way, thus yielding more conservative posterior inferences. 

\par In Bayesian stacking, each candidate model in the ensemble is fitted to the observed data to permit the estimation of the model parameters, or functions of these, allowing for an improved predictive accuracy.    In our paper we focus on combining Bayesian stacking with a related application of \emph{posterior predictive checking} (PPC).
Based on Bayesian stacking, the predictions from the candidate models are combined to additionally reflect model uncertainty, by basing predictions on a weighted combination of the candidate models, where the weights are obtained using the expected log predictive density based on leave-on-out cross-validation.
Using PPC, a special type of model diagnostic in Bayesian statistics, we refine our understanding of the results and associated underlying factors from a certain perspective in terms of content.
More precisely we use PPC to assess the validity of the stacking ensemble, thus the stacked predictive
distributions of the candidate models, with respect to our substantive research question and to learn from deficiencies of the candidate models with regard to predicting certain regions of the outcome distribution. In fact, PPC  is an integral part of the process for assessing the extent to which a certain model satisfies aspects of the research question in specific sub-populations. The underlying idea of PPC is to benchmark actual observations to a simulated distribution of hypothetical realizations under a specific model, which allows for criticizing the model fit \citep*[see e.g. ][and references therin]{kaplanbayesbook2,  gelman2013bayesian, vandeSchoot.2021}\footnote{We distinguish between the method of as a form of external validation.\textit{posterior predictive checking} (PPC) and posterior predictive checks (PPCs) as their simulation-based realizations.}.  In our workflow, we perform checks within levels of grouping variables. 

These PPCs also allow us to explore the variation of the predictions based on the different candidate models we compare during the model building process.  

\subsection{Real-world Example and Structure of the Paper}

\par In this paper, we apply Bayesian stacking and PPC to data from PISA 2018 \citep{pisa.2020} to investigate potential inequalities in reading competency with respect to gender and socio-economic background. Reading competency is a heavily discussed topic in educational research, and in particular in sociology of education, and several explanatory concepts exist, which lead to potentially very different modeling approaches.

We deliberately selected this example, because of its relevance to research in social sciences \citep*[see e.g.][]{ine, Coleman1993, Jacobs, Buchmann, Becker.2018}, as we want to demonstrate how the proposed approaches can help to gain new insights for existing research questions. In particular, we will emphasize several key advantages of our approach.
 
\par The paper continues with an overview of predictive modeling from a Bayesian perspective, and moves on to discussing model uncertainty via Bayesian stacking and the basic idea of PPC in sections \ref{sec:PredMod} and \ref{sec:stack}. After that we will introduce the PISA 2018 data in section \ref{sec:EmpAppl}, as well as all relevant aspects for our empirical example. Section \ref{sec:Workflow} contains a description of the workflow as a rough guideline for practitioners who want to implement the concepts of predictive modeling and model uncertainty in their work. In section \ref{sec:extensions} we mention some possible extensions of the workflow. We summarize the findings of our work and discuss the aforementioned key advantages in section \ref{sec:Outlook}.

%%%%%%%%%%%%%%%%%%%%%%%%%%%%%%%%%%%%%%%%%%%%%%%%%%%%%%%%%%%%%%%%%
%%%%%%%%%%%%%%%%%%%%%%%%%%%%%%%%%%%%%%%%%%%%%%%%%%%%%%%%%%%%%%%%%

\section{Model Uncertainty and Predictive Modeling under a Bayesian Framework} \label{sec:PredMod}

\par A major argument made in this paper is that a fundamental goal of statistics is the development of predictive models with optimal predictive performance.  All other things being equal, a given model is to be preferred over other possible models if it provides better predictions of what is actually observed \citep{Dawid84}. In fact, it is hard to feel confident about inferences drawn from a model that performs badly in predicting the observed data. That said, the difficult question is how to develop accurate predictive models, and, especially, how to assess their accuracy. The approach taken in this paper is \emph{Bayesian stacking} which is a special case of \emph{model ensembling} \citep{Wolpert1992,LBreiman.1996,YulingYao.2018}. 
\par We begin this section with an introduction to Bayesian predictive modeling.  In particular we review the method of Bayesian model averaging briefly. We then outline a major assumption underlying Bayesian model averaging which can limit its usefulness in the social sciences.  We then, in section \ref{sec:stack}, move to Bayesian stacking as alternative approach that addresses the limitations of Bayesian model averaging and focuses on combining the predictive distributions from user-specified models. Finally we discuss PPC as a tool for and its evaluation in Bayesian predictive modeling.

\subsection{BMA and its Limitations}

\par In the context of model uncertainty, the problem that arises is that the selection of a single model from a class of models that is then used for prediction or inference ignores the information contained in the other models. This leads to ``...over-confident inferences and decisions that are more risky than one thinks they are" \citep[][pg. 382]{hoeting99}.  A major contribution to addressing this problem had been the development and application of \emph{Bayesian model Averaging} (BMA) \citep*{clyde99,Clyde2003,draper95,leamer,madiganraftery94,raftery97,hoeting99}.  
Simply put, BMA considers models to be elements of a discrete space of models described by a probability mass function.  Choosing a single model and ignoring the information in the remaining models implies that 100\% of the probability mass lies on that model - or in other words, the chosen model is assumed to be the model that generated the data. By contrast, BMA enables an exploration of the model space, where first it is usually assumed that each model is equally likely to be the true data-generating model. This assumption is encoded into BMA by specifying a non-informative probability mass function over the space of possible models.  Then, each model is confronted by the data, and, by way of Bayes' theorem, this leads to posterior model probabilities that will vary across the model space because not all models are equally good in the sense of posterior model probability.  Rarely in practice will one find a single model with posterior probability 1.0, and indeed, the "best" model will often have a posterior model probability considerably less than 1.0, reflecting the extent of model uncertainty. 

\subsection{$\mathcal{M}$-Frameworks}
\par Despite its popularity and success across a wide range of content domains \citep*[e.g.][]{fernandezleysteel01b, MontgomeryNyhan2010, yeungbumgarnerraftery, Raftery2005UsingBMA,sloughteretal,kaplanchenbma,kaplanleebmasem, miBMA, KaplanHuang}, a critical assumption underlying BMA is that the true data generating model, denoted as $M_T$, is one amongst finitely many models explored by the BMA algorithm. If we make the assumption that $M_T$ is in the  model space, this is referred to as the $\mathcal{M}-{closed}$ framework, introduced by \citep{bernardosmith2000} and further discussed in \citep{ClydeIversen2013}. 
 As long as it is possible for the researcher to assign model priors the the $\mathcal{M}-{closed}$ framework can be adopted.
However, the truth or falsity of the
$\mathcal{M}-{closed}$ framework notwithstanding, it is essential to reiterate that conventional BMA assumes the $\mathcal{M}-{closed}$ framework and, indeed, readily available BMA software typically employ a non-informative prior to the space of models as a default, with the idea that the true model is in the model space. 
In distinction,
$M-{complete}$  frameworks assume that a true model exists, but is intractable in the sense that it cannot be
 used directly in the analysis. In practical terms, this implies that
 the true model is not within the set of models under consideration and therefore other
surrogate models must be specified and used in the analysis.
If it is difficult to justify model priors as required under $M-{closed}$, and if model selection based on maximizing the expected utility under $M-{complete}$  is not convincing, then  approach that allows for model averaging without the need to assume that $M_T \in \mathcal{M}$ is required. This is referred to as the  $\mathcal{M}-{open}$ framework \citep{bernardosmith2000}.The next sections describes model averaging in the $\mathcal{M}-{open}$ framework, namely  the methodology of \textit{Bayesian stacking}.

%%%%%%%%%%%%%%%%%%%%%%%%%%%%%%%%%%%%%%%%%%%%%%%%%%%%%%%%%%%%%%%%%
%%%%%%%%%%%%%%%%%%%%%%%%%%%%%%%%%%%%%%%%%%%%%%%%%%%%%%%%%%%%%%%%%

\section{Theoretical Background of Stacking and Performance of Predictive Models} \label{sec:stack}

\subsection{Theoretical Background of Stacking}

\par Following closely the review by \citet{Kaplan.2021}, the method of \textit{stacking} was originally developed in the machine learning literature by \citet{Wolpert1992} and \citet{LBreiman.1996} and brought into the Bayesian paradigm by \citet{ClydeIversen2013}.  Using Bayesian stacking the analyst enumerates a set of $K$   $(k=1,2,\ldots K)$ models and then creates a weighted combination of their predictions. Although the initial purpose of the method was to stabilize predictions and, thus, minimize their mean squared error, Bayesian stacking can also be used to account for model uncertainty by focusing on the differences in predictive distributions obtained from different candidate models.  For instance, we can specify several candidate models of reading competency as 
$f_{k}$,
where $f_{k}$ refers to different models of the outcome.  Next posterior predictions from these models are combined using their scores as weights as \citep[see][]{LeClark2017}
\begin{equation}
\tilde{y} = \sum_{k=1}^K \hat{w}_{k}\hat{f}_k(x),
\end{equation}
where $\hat{f}_k$ estimates $f_k$.  The weights, $\hat{w}_k$ $(\hat{w}_1,\hat{w}_2,\ldots \hat{w}_K)$ are obtained as
\begin{equation}\label{stackingweight}
\hat{w} = \argmin_w\sum_{i=1}^n\left(y_i - \sum_{k=1}^K w_{k}\hat{f}_{k,-i}(x_i)\right)^2
\end{equation}
where $\hat{f}_{k,-i}(x_i)$ is an estimate of $f_k$ based on $n-1$ observations, leaving the $i^{th}$ observation out.This variant of k-fold cross validation is called \textit{leave-one-out cross validation} (LOOCV), and it ensures that the investigated observation does not impact its own predictive distribution. 
LOOCV is available in the
\textsf{R} package \textsf{loo}
\citep*{loo_R}. For further details regarding LOOCV we refer readers to \citet{Kaplan.2021}.

\subsection{Evaluating the Performance of Predictive Models}
\par As noted earlier, the motivating reason for model stacking is to obtain a model with optimal predictive performance.  Among a set of different methods such as BMA, stacking, or regularization (which does not account for model uncertainty)\footnote{Regularization methods such as the ridge, lasso or horseshoe prior indeed address the bias-variance trade-off and can lead to acceptable predictive performance, they do not explicitly address model uncertainty insofar as one model is ultimately used for prediction and inference \citep{kaplanbayesbook2}}, a reasonable method for evaluation should be based on measures that assess predictive performance -- referred to as \emph{scoring rules}.  

%\subsubsection{\textbf{Strictly Proper Scoring Rules}} 

\par We apply scoring rules in order to 
evaluate the performance of predictive models. A prediction is said to be ``well-calibrated'' if the assigned probability of the outcome match the actual proportion of times that the outcome occurred \citep{dawid82}. A large number of scoring rules have been discussed in the literature \citep*[see e.g.][]{Winkler96,bernardosmith2000,JoseNauWinkler08,merklesteyvers,gneitingraftery}. In this paper we use the Brier score which is used for dichotomous outcomes.
 The Brier score can be defined as
\begin{equation}
-(||\mathbf{e}_i - \mathbf{p}||_2)^2
\end{equation}
where $\textbf{e}$ denotes an indicator vector of an event that is estimated by its forecast $\textbf{p}$. For example, $\textbf{p}$ may represent the forecast probability of being a good reader, and $\textbf{e}_i$ represents the realization of the event scored 1/0.  The Brier score penalizes the forecaster in proportion to the squared Euclidean distance between the forecast and the event \citep[1148]{JoseNauWinkler08}

\subsection{Posterior Predictive Checking}

\par In comparison to the standard research cycle, the Bayesian workflow contains a form of  model criticism, namely PPC, which aims to understand
to what extent the models fit the data reasonably well regarding the research question of interest, and to what extent they are inadequate \citep*[see e. g.][]{bayesWF}. Thus, one uses  PPC  to ensure that the model is appropriate to describe the observed data for the given research interest by searching for systematic discrepancies between the fit of the model and the data. 
  
\subsubsection{Background of Posterior Predictive Checking}

\par The basic idea is to
simulate data under the fitted model, which here encompasses the complete model specification, and to compare summary statistics of the replicated data  to the corresponding summary statistics of the observed data  \citep[see e.g.][]{kaplanbayesbook2,  gelman2013bayesian, vandeSchoot.2021}. A key advantage of PPC is  that  predictive simulation explicitly accounts for the inferential  uncertainty that is usually neglected by standard approaches \citep{LynchWestern}.  A second benefit is that a large number of test statistics can be defined to check for specific inconsistencies between the observed data and the model predictions. However, PPC makes use of the data twice and therefore the researcher must define test statistics  carefully in order to ensure that  PPC is able to detect conflicts between model and data \citep*{Gabry_2019}. 
 Thus the construction of the test statistics and interpretation of the corresponding PPCs force the researcher to question what content aspects are actually represented by the model. If the observed data are reasonable under the predictive distribution of the replicated summary statistics, the model  has "passed" the check. 
 Graphical PPC means to create graphical displays comparing observed data to simulated data from the posterior predictive distribution \citep{Gabry_2019}.
 Bayesian  statistics allows the calculation of from of $p$-value: The posterior predictive $p$-value, also referred to as Bayesian $p$ value, is defined as the probability that the replicated
data could be more extreme than the observed data, as measured by the test quantity \citep{gelman2013bayesian}. 
The posterior predictive distribution is defined as $\mathbb{E}_{\boldsymbol{\zeta}|\boldsymbol{y}} \bigl[ p(\boldsymbol{y}^{\text{rep}}|\boldsymbol{\zeta}) \bigl],$ 
thus
    $$
    p(\boldsymbol{y}^{\text{rep}}|\boldsymbol{y})=
\int p(\boldsymbol{y}^{\text{rep}}, \boldsymbol{\zeta} |\boldsymbol{y}) \: d \boldsymbol{\zeta}  = \int p(\boldsymbol{y}^{\text{rep}}| \boldsymbol{\zeta}, \boldsymbol{y}) p(\boldsymbol{\zeta}| y) \: d \boldsymbol{\zeta} 
    $$
    \begin{equation}\label{PPD}
= \int p(\boldsymbol{y}^{\text{rep}}| \boldsymbol{\zeta}) p(\boldsymbol{\zeta}| y)  \: d \boldsymbol{\zeta},
\end{equation}

 where $\boldsymbol{y}^{\text{rep}}$ is the replicated data set,
 $\boldsymbol{y}$ is the observed data set and $\boldsymbol{\zeta}$ is the parameter vector.
\par Based on draws of the posterior predictive distribution one can compute certain test statistics $T$ like e. g. the mean, quartiles or the standard deviation. In this way the
distribution of $T(\boldsymbol{y}^{\text{rep}})$ is derived and
serves as
reference distribution for $T(\boldsymbol{y}).$ The Bayesian $p$-value
is a numerical measure to summarize the comparison of $T(\boldsymbol{y})$
to that reference distribution.
The one-sided posterior predictive $p$-value for a certain statistic test $T$  is:
    \begin{equation}
        \mathbb{P} \bigl( T(\boldsymbol{y}^{\text{rep}}) >  T (\boldsymbol{y})| \boldsymbol{y}\bigl) = \int_{[T(\boldsymbol{y}^{\text{rep}}):T(\boldsymbol{y}^{\text{rep}}) >  T(\boldsymbol{y})]} p(\boldsymbol{y}^{\text{rep}}|\boldsymbol{y})\: d \boldsymbol{y}^{\text{rep}}.
    \end{equation} 

This Bayesian $p$-value measures the proportion of test statistics $T(\boldsymbol{y}^{\text{rep}})$
that is greater than the test statistics based on the actual data $T(\boldsymbol{y})$.
If, for substantive reasons, a deviation between the replicated data
 and the actual data implies model misspecification, than values
 of the one-sided Bayesian $p$-value 
closer to $0$ or $1$ indicate a poor posterior predictive qualities \citep{kaplanbayesbook2}.

\par If either small values of the test statistic $T(\boldsymbol{y})$ or high values $T(\boldsymbol{y}),$
for substantive reasons, indicate a poor model fit, 
the two one-sided Bayesian $p$-values is to be used. Per definition the two sided posterior predictive $p$-value is twice the minimum of the two one-sided
Bayesian $p$-values.
Thus, the two sided posterior predictive $p$-value 
equals $1.0$ if the observed test statistic is equivalent to the median
of the distribution of the replicated test statistic. Values below $1.0$ indicate that posterior prediction of the model might not be as reliable as it should be. Determining the seriousness of the problem is a matter of substantive judgement
and therefore we deliberately do not define a cutoff to determine if a model has passed a certain check. The two sided posterior predictive $p$-value was calculated with the R package \texttt{HuraultMisc} \citep*{Hurault.2021}, and is denoted \textit{TSPPPV} in the following. 

\subsubsection{Application of Posterior Predictive Checking in this Work} \label{appppc}

We aim to find a model with high predictive quality that predicts the outcome given a list of  predictors. We use PPC to check whether certain regions of the outcome distribution are predicted accurately. In order to achieve this goal, we perform PPCs within levels of grouping variables to explore in how far the candidate model and the stacking ensemble predict these regions of the outcome distribution. Therefore, both the outcome and the predictive distribution is split based on one or more grouping variables to compare the observed group-specific test statistic $T(\boldsymbol{y})$ (which is the mean in our applied example) to the corresponding reference distribution.
Both graphical PPC and the posterior predictive $p$-value are used to detect systematic discrepancies
between model fit and observed data in certain regions of the outcome distribution.

\par In fact, we use PPC for two different purposes: In Section \ref{step41}, we examine whether regions of the outcome distribution that are directly related to the primary substantive research question are correctly predicted by the stacking ensemble.  If not, one must take this into account when interpreting the results. In this case the PPCs are used exclusively to check the stacking ensembles.
 In section \ref{step42}, PPC is being used to investigate the validity of the stacking ensemble. The stacking ensemble is to be used at the end of the analysis   to derive model-based predictions for values of the predictors being relevant in terms of content, which represent typical students. The aim is to develop a stacking ensemble that makes optimal predictions for those virtual 'representative' students. In general, a statistical model is valid if it aligns with the research question \citep[see][153]{Gelman.2021}. This implies that
the outcome reflects the phenomenon of interest for all values of the predictors, which are relevant in terms of content, and it implies in particular that the model contains all relevant predictors. The model should universally fit the statistical units to which it is applied.
In order to test this we investigate a predictor which we deliberately omitted from the model: The variable \textit{WITHOUT}, which equals $1$ if the school percentage of students who leave without certification is minimum  2\%, and $0$ otherwise. An indication that the stacking ensemble is
beneficial is the correct prediction of the corresponding regions of the outcome distribution.
We perform the PPCs for the candidate models as well as the stacking ensemble, because it helps us to investigate and understand the validity of the stacking ensemble. 

%%%%%%%%%%%%%%%%%%%%%%%%%%%%%%%%%%%%%%%%%%%%%%%%%%%%%%%%%%%%%%%%%
%%%%%%%%%%%%%%%%%%%%%%%%%%%%%%%%%%%%%%%%%%%%%%%%%%%%%%%%%%%%%%%%%

\section{Empirical Application}

\label{sec:EmpAppl}
\subsection{Overview of the PISA Study} 

The Program for International Student Assessment (PISA) was launched in 2000 by the Organization for Economic Cooperation and Development (OECD) as a triennial international survey designed to evaluate education systems worldwide by testing the skills and knowledge of 15-year-old students.  In 2018, 600,000 students, statistically representative of 32 million 15-year-old students in 79 countries and economies, sat for an internationally agreed-upon two-hour test. The subjects in which the students were assessed are science, mathematics, reading, collaborative problem solving, and financial literacy.From the context of country-level policy relevance, PISA is the most important international survey that is currently operating \citep{pisa2000techreport}.
\par The sampling framework for PISA follows a two-stage stratified sample design \citep{KaplanKuger2016}.  Each country/economy supplies a list of all ``PISA-eligible'' schools.  This list constitutes the sampling frame. In a next step schools are sampled from this frame. The sampling probabilities are proportional to the size of the school, with the size being a function of the estimated number of PISA-eligible students in the school.  The second-stage of the design samples students within the sampled schools.  A target cluster size of 35 students within schools was  targeted, however for some countries, this desired cluster size was negotiable.  The spiraling design and plausible value methodologies that are used in the method of assessment for PISA follow closely the design for the U.S. National Assessment for Educational Progress (NAEP) \citep[see][]{NAEP, pisa2015techreport}

\par In addition to these so-called ``cognitive outcomes", policymakers and researchers alike also focus increasing attention on the non-academic contextual aspects of schooling. Context questionnaires provide important variables for models predicting cognitive outcomes and these variables have become important outcomes in their own right - often referred to as ``non-cognitive outcomes" \citep[see e.g.][]{HeckmanKautz2012}. PISA also assesses these non-cognitive outcomes via a one-half hour internationally agreed-upon context questionnaire \citep*[see][]{KugerJudeKliemeKaplan2016}.
Data from PISA is freely available and can be accessed via their \emph{PISA Data Explorer} at \url{https://www.oecd.org/pisa/}.

\subsection{\textbf{PISA Survey Weighting}}
\par The nature of the sampling design for PISA ensures that the sample of students for a given country is chosen randomly and thus reflects the national population of 15 year olds for that country.  However, within countries, the selection probabilities to attain national representativeness might be different and so survey weights should be used to ensure that each sampled student represents the appropriate number of students in the PISA-eligible population within a particular country.  

\subsection{\textbf{PISA Scaling}}
\par A three-step process is used to generate test scores in PISA and this process has been quite consistent over the cycles of PISA.  These steps include (a) obtaining national calibrations, (b) performing an international calibration, and (c) generating distributions of student proficiency.  The overarching methodology across these three steps is \textit{item response theory} (IRT) \citep{lordnovick}; a method for relating item responses on some measure to an underlying latent trait, such as mathematics proficiency through a probabilistic model.

%%%%%%%%%%%%%%%%%%%%%%%%%%%%%%%%%%%%%%%%%%%%%%%%%%%%%%%%%%%%%%%%%
%%%%%%%%%%%%%%%%%%%%%%%%%%%%%%%%%%%%%%%%%%%%%%%%%%%%%%%%%%%%%%%%%

 \section{Workflow for Predictive Modeling Using  Bayesian Stacking and Posterior Predictive Checking} \label{sec:Workflow} 
 
The contents of this section can be considered as a template for using Bayesian predictive modeling in combination with Bayesian stacking and PPC, which is why we also explicitly mention the most relevant functions and packages in \textsf{R} which were used for the implementation. 

Specifically, we use an applied example to demonstrate the potential of Bayesian stacking for assessing social inequalities between particular groups based on predictive modeling, while simultaneously accounting for model uncertainty.  The empirical example serves as a rough guideline for practitioners who want to implement Bayesian stacking and posterior predictive checking in their own work.
Our research workflow comprises the stacking of logistic regression models to analyze which groups of 15-year-old students at the end of their compulsory education are at risk of not achieving the minimum proficiency in reading.
Throughout the analysis we denote students who performed below the minimum reading proficiency level, i.e. below proficiency level 2, as \textit{literacy deprived students}. 
  The workflow consists of five key steps:
  \begin{enumerate}
      \item Defining the research question and identifying regions of the outcome distribution of substantive interest
      \item Specifying candidate models 
      \item Estimating models using Bayesian stacking and MCMC diagnostics
      \item Posterior predictive checking
      \item Summarizing results based on the posterior distribution
  \end{enumerate}
  
 \par In the following we address the research question on a methodological level. 
 The example only serves to illustrate a possible workflow and the results have no significance in terms of content. In particular, we do not address necessary adjustments to account for the complex survey and test design of PISA. However, the statistical approach presented here can be used for research on similar methodological questions.

\par Data from international large-scale assessment studies underline ongoing inequalities with respect to reading competency, in particular related to gender and/or socio-economic background (depending on country), which is often measured based on the index of economic, social, and cultural status (ESCS). The PISA index ESCS is derived from three variables related to family background: parents’ highest level of education ,parents’ highest occupational status, and home possessions, including books in the home \citep{pisa.2020}. 
 In general, girls outperform boys and socio-economically advantaged students outperform disadvantaged students. However, the extent of the gender gap can be highly dependent
on ESCS \citep[see ][Chap. 7]{pisa2018p}.  
The simultaneous consideration of gender and socio-economic background is of substantive interest to understand and explain the development of social inequalities.

 \par We picked Greece as an example, however the analysis could be performed in a similar way for other countries as well. The gender gap for reading competency in Greece is above average in comparison to the OECD mean. In general, Greece scored lower than the OECD average and the situation has become even worse since 2009 \citep[134 ff]{pisa.2020}, potentially due to cuts in the education system \citep{Liaropouos.2020}. Under these circumstances, a nuanced description of inequalities that is transparent with respect to statistical assumptions is important. The aim of our analysis is to demonstrate a differentiated description of social phenomena, providing an empirical basis for a substantive discussion of the social phenomenon  by capturing underlying complex patterns and relationships between key predictors. A major argument made in this paper is that we incorporate model uncertainty in an intuitive and coherent way. The social phenomenon we are interested in is the gender gap in reading, taking into account the socio-economic background. Thus, our primary research interest in this example is the differential representation of the gender gap in reading. 

\subsection{Step 1: Research Interest and Initial Variable Selection}

\par In the following, we use the term "variable" for the variables in the PISA data set, and we refer to the summands in the linear predictor of a model as "predictors". 
The outcome variable in our example is the binary predictor of being literacy deprived in reading\footnote{We only use
the first plausible value of the corresponding variable in the PISA data set to define literacy deprivation.}. 
 Literacy deprived students are students who performed below reading competency level 2. Throughout our analysis we particularly focus on subgroups defined by gender as predictor (binary coded\footnote{If gender is surveyed in three categories, the measured gender gap would refer to female versus non-female.}) and socio-economic background, measured in quartiles of ESCS and implemented as predictor variable \textit{SES}\footnote{Note that the variable \textit{SES} results in three predictors, namely \textit{SES2}, \textit{SES3} and \textit{SES4}, 1 = bottom quarter, 2 = second quarter, 3 = third quarter, 4 = top quarter.
    The bottom quartile is used as base level.}.  

\begin{table}[H]
    \caption{Overview of model Variables: Outcome \textit{LD} and  two focal variables. \vspace{0.2 cm}}
    \centering
    \begin{tabular}{|c|l|}
    \hline
    Variable label & Description \\
    \hline
    \textit{LD} & Literary deprived student (binary, 1 if yes) \\   \hline
    \textit{FEM}        & Gender (binary, 1 if reported female)\\  \hline
    \textit{SES}        & Quartiles of index of economic, social, and cultural  status (ESCS)\\ \hline
\end{tabular}
 \label{tab:Varoverview}
\end{table}

 \par 

 \subsection{Step 2: Specification of the Core Model and the Candidate Models}

 \par We follow the recommendation by \citet[][199]{Gelman.2021}for building predictive regression models, to specify candidate models and include all variables that, for substantive reasons, might be expected to be important in predicting the outcome. We use the constructed indices as suggested by \citet{pisa.2020} to further reduce the size of the vector containing potential model variables. The final list consists of $40$ potentially relevant  variables of the PISA data set. Additionally, we include $24$ interaction terms to further improve 
the predictive accuracy of the model, the final list consists of $66$ predictors.
\par This step resembles the phase of the 'normal' model building process, where we would try out several models. We deviate from it, however, as we are not going to select a single 'winner' among the candidate models. Since we assume $\mathcal{M}-{open}$ we do not try to represent the data generating process by these models, but to specify models with high prediction performance.

 \par Candidate models ideally have a high predictive accuracy for different regions of the outcome distribution. We sort the list of initial variables such that each candidate model includes a certain set of variables important to describe the 
heterogeneity in the distribution of the outcome. Variables are sorted according to various factors:
\begin{itemize}
    \item Candidate model 1 comprises variables associated with student attitudes and dispositions towards reading (\textit{reading attitude model})
    \item Candidate model 2 comprises variables related to the extent of teachers improving reading competency as well as attitude towards school or learning  (\textit{learning attitude model})
    \item Candidate model 3 comprises modifiable variables on the school level, e.g. classroom composition (\textit{composition model})
    \item Candidate model 4 comprises descriptive (non-modifiable) variables on the school level, e.g. classroom discipline (\textit{school climate model})
\end{itemize}

The distinction between variables in models 3 and 4 follows a suggestion by \citet*{JEHANGIR20151}.
If we assume that it improves predictive accuracy, the corresponding interaction term is additionally included in the model specification\footnote{The appendix contains the detailed table with all variables and interactions included in the candidate models, see \ref{app:step2} on page \pageref{app:step2})}. 

 \par The number of candidate models and their respective functional form  depends on the prior knowledge that a researcher has about the predictive performance of particular variables and on the region of the outcome distribution
 the researcher wants to address. In  particular, interactions between focal variables should be included in the model specification because the model thereby obtains flexibility for predicting the corresponding regions of the outcome distribution.
 It is important to emphasize that all models that were specified are actually included in the analysis by using Bayesian stacking. Bayesian stacking then combines these candidate models by weighting them in order to produce a stacked predictive distributions of the candidate models, that has the highest overall predictive quality.
 
 \par The variables \textit{FEM} and \textit{SES} given in table \ref{tab:Varoverview} are included in each candidate model, because the corresponding regions of the outcome distribution are of key interest. The related  model is denoted as the \emph{core model}.
The fact that the two variables of the core model are present in all candidate models makes it possible to take into account corresponding interactions with these predictors and other variables.
Further variables included in the candidate models serve to improve the predictive accuracy of the stacking ensemble. The candidate models share the core model and are disjoint regarding the other variables.

\subsection{Step 3: Model Estimation Using Bayesian Stacking and MCMC Diagnostics of Candidate Models}

 \par Since the posterior and posterior predictive distributions are typically analytically intractable, Bayesian statistics relies on simulation methods for approximating the target density. We used the \proglang{rstanarm} package in \proglang{R} \citep*{rstanarm} for our regression models which, in turn, uses the \proglang{Stan} sampling engine \citep{StanReferenceManual} to generate simulation draws based on the No-U-Turn implementation of Hamiltonian Monte Carlo (HMC) \citep{HoffmanGelman2014}. HMC is a particular variant of Markov Chain Monte Carlo (MCMC) procedures, and we will use the term 'MCMC' throughout this paper, by which we mean HMC as implemented in \proglang{Stan}. Since the (usually arbitrarily chosen) starting values to initiate the (Markov) chains can influence the values of high-order lags, the first draws are discarded as 'burn-in' and the researcher needs to check for convergence in distribution. MCMC diagnostics are tools used to verify that the quality of a sample generated by a MCMC algorithm is sufficient to provide an accurate approximation of the target distribution, and detailed information about these MCMC diagnostics can be found in \citet{StanReferenceManual}.
After checking convergence the researcher summarizes the regression results based on 
the posterior distribution of the regression coefficients\footnote{see appendix \ref{regRes} on page \pageref{regRes}}. 

 \par Within this step, the Brier score is used for the predictive accuracy assessment of the core model, the stacking ensemble and the candidate models, where the stacking ensemble is expected to have the best predictive accuracy. Table \ref{tab:wtsBrier} informs about the stacking weights of the core  model and the four candidate models. Candidate model 2 has the highest stacking weight.
 Obviously, the stacking ensemble has the lowest Brier score.

\begin{table} [h!]
   \caption{Stacking weights and Brier scores } \vspace{0.2cm}
   \centering
   \begin{tabular}{rrr}
\hline
    Model & Stacking weight & Brier score \\ 
    \hline
    core m. & / & 0.183 \\ 
   % \hline
    stacking e. & / & 0.138 \\ 
  %  \hline
    reading attitude m. (cand 1) & 0.100   & 0.158 \\ 
  % \hline
     learning attitude m. (cand 2) & 0.533   & 0.144 \\
 %  \hline
    composition m. (cand 3) & 0.172  & 0.151\\ 
%    \hline
     school climate m. (cand 4) & 0.195 & 0.151 \\ 
    \hline
\end{tabular}
\label{tab:wtsBrier}
\end{table}

\subsection{Step 4: Posterior Predictive Checks}
The first part of the post-modeling phase  consists of using PPC to examine the predictive quality of those aspects of the stacking ensemble that are important to the research question.
We use the mean as a test statistic $T(\boldsymbol{y})$ for PPCs to compute the group-specific prevalence of being
literacy deprived. 

   \subsubsection{Posterior Predictive Checks for the Focal Predictors}
\label{step41}

\begin{figure}[h]
 \centering
    \includegraphics[width=0.4\linewidth,page=1]{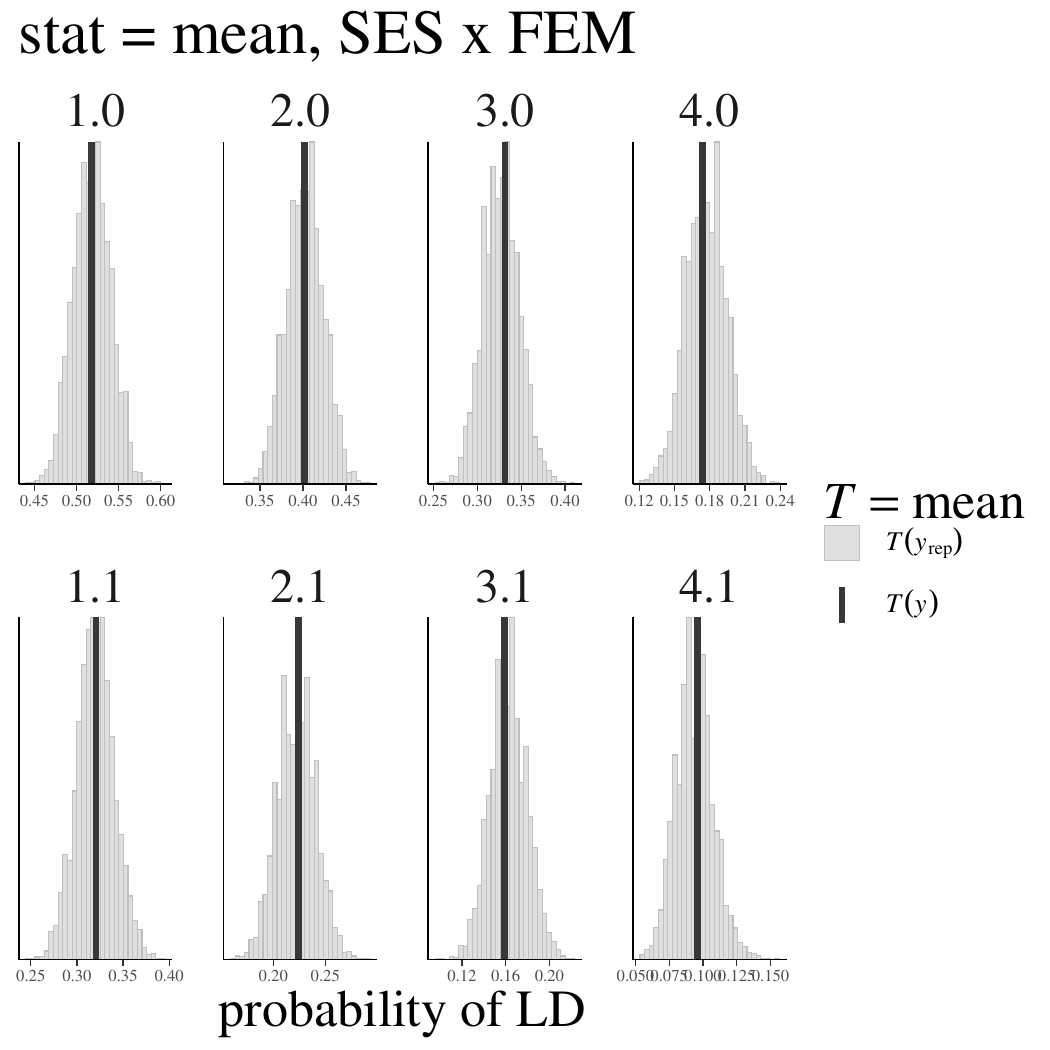}
 \caption{PPCs based on 4000 simulation draws within levels of the grouping variables \textit{SES} x \textit{FEM}: The top and bottom row display the four \textit{SES}-quartiles for male and female students, respectively. \label{stack_in}}
\end{figure}

 We consider the posterior predictive distribution of being literacy deprived (LD) based on the PPC simulation draws for pre-defined subgroups, and we start out by using the focal variables given in table \ref{tab:Varoverview} as grouping variables. The corresponding PPCs are helpful for assessing if the stacking ensemble is appropriate for our research question. The graphical results for the combinations are displayed in figure \ref{stack_in}. The actually observed value for the prevalence of literacy deprived  students is represented by the solid black line, and the posterior predictive distribution based on simulation draws is represented as histogram. The observed prevalence is in all cases very close to median, which indicates a very good fit of the stacking ensemble.  The corresponding two sided posterior predictive $p$-values (TSPPPV) are given in table \ref{tstack_in}.  Ideally, the TSPPPV is $1.0$, whereas values less than $1.0$ indicate discrepancies between the data and the model.

\begin{table}[!htb]
    \caption{Two sided posterior predictive $p$-values (TSPPPVs) for combinations of \textit{SES} x \textit{FEM} that correspond to the graphical PPCs in figure \ref{stack_in}.  \label{tstack_in}}
    \centering
      \begin{tabular}{|r|r|r|r|r|}
  \hline
& \textit{SES} = 1 & \textit{SES} = 2 & \textit{SES} = 3 & \textit{SES} = 4\\     
 \hline
\textit{FEM = 0} & 1.00 & 0.98 & 0.80 & 0.91\\
\hline
\textit{FEM = 1} & 0.95 & 0.91 & 0.85 & 0.82\\
\hline
\end{tabular}
\end{table}

\subsubsection{Posterior Predictive Checks to Investigate and Understand the Validity of the Stacking Ensemble}
\label{step42}

\par The four candidate models each contain a list of variables that represent different facets of explanations for literacy deprivation, and we therefore refer to them as the \textit{reading attitude model}, the \textit{learning attitude model}, the \textit{composition model} and the \textit{school climate model}. These  PPCs help us to explore the extent to which the respective models succeed in describing content-specific regions of the outcome distribution. We first compare the core model and the stacking ensemble, and we subsequently investigate the four candidate models. 

\begin{figure}[h]
  \begin{subfigure}{.33\textwidth}
  \centering
    \includegraphics[width=1\linewidth,page=2]{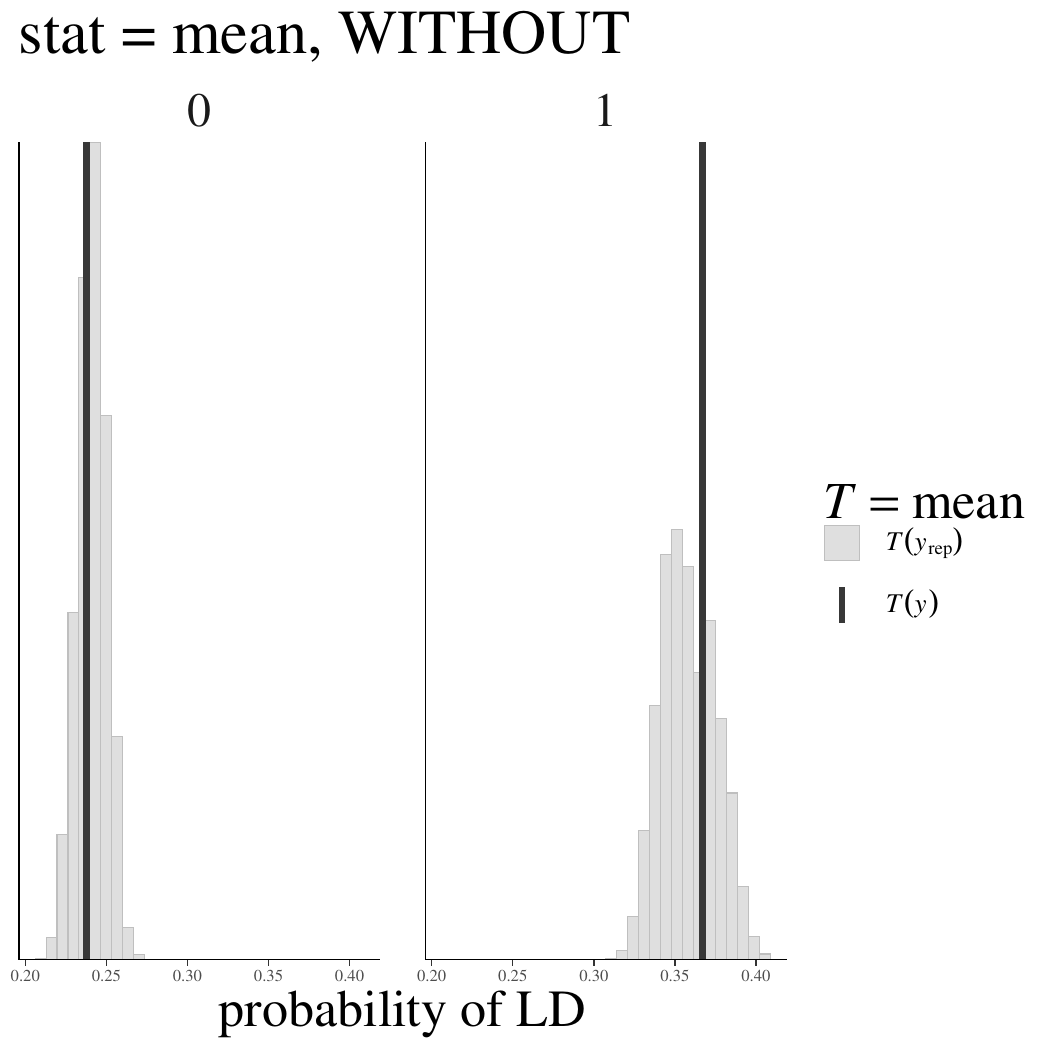}
    \caption{Stacking ensemble}
  \end{subfigure}%
  \begin{subfigure}{.33\textwidth}
  \centering
    \includegraphics[width=1\linewidth,page=2]{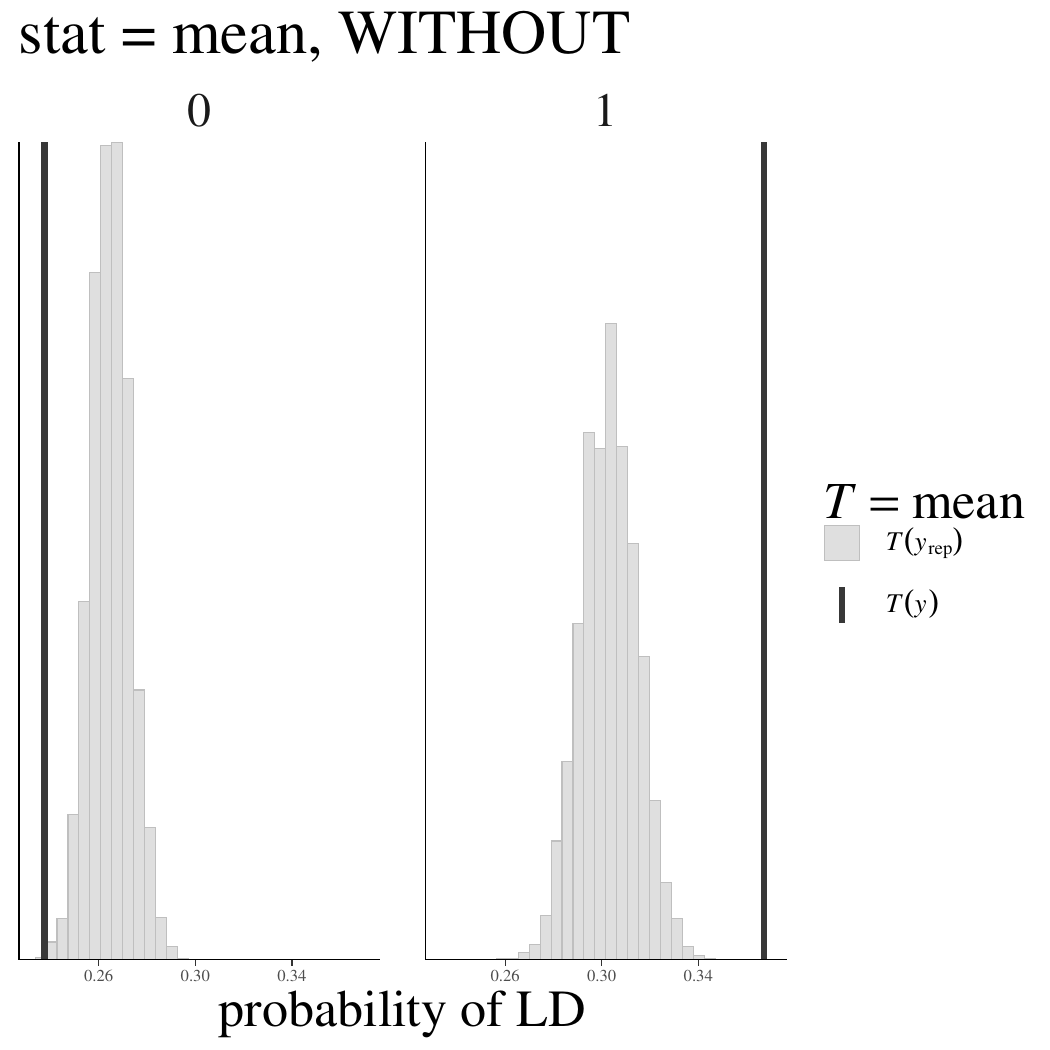}
    \caption{Core model}
  \end{subfigure}
  \begin{subfigure}{.33\textwidth}
  \centering
    \includegraphics[width=1\linewidth,page=2]{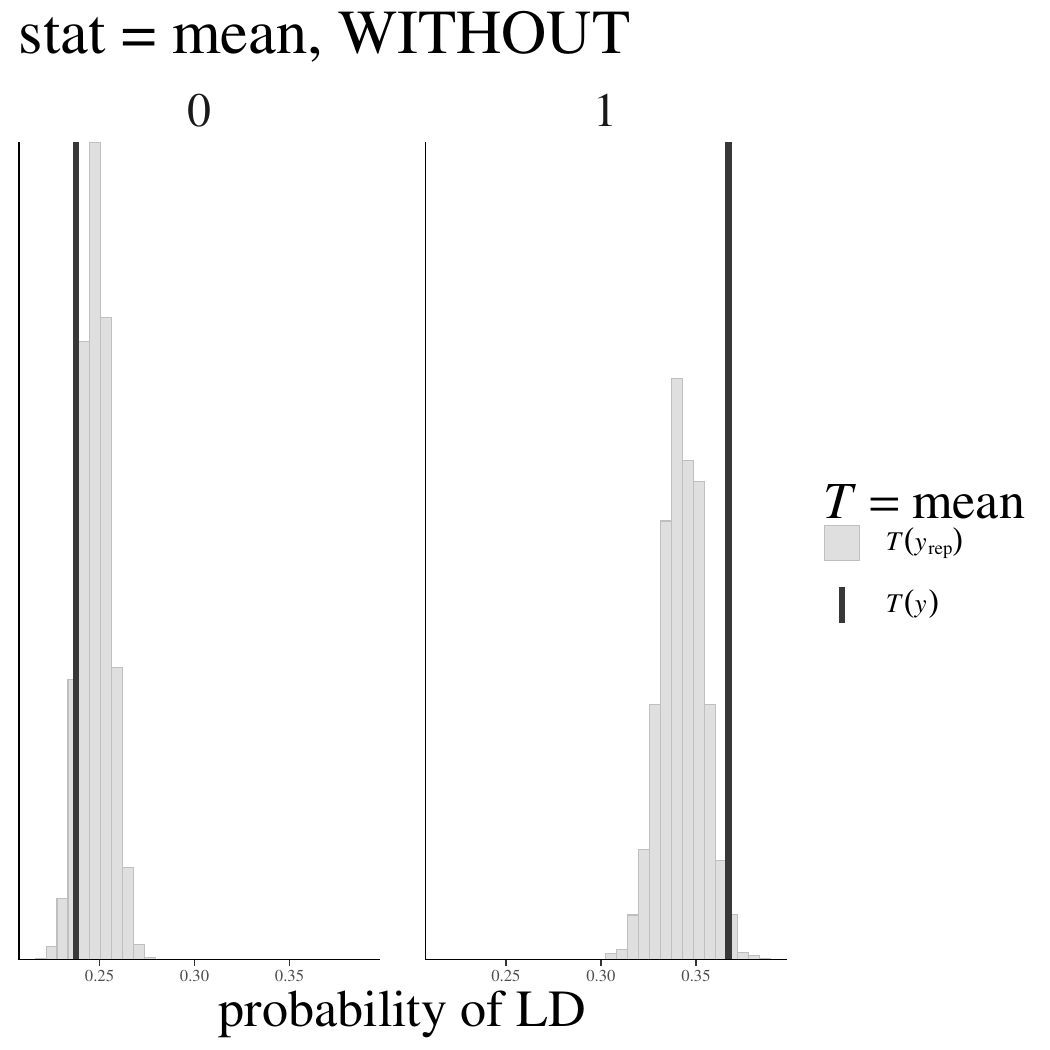}
    \caption{Candidate model 1}
    \end{subfigure}
      \begin{subfigure}{.33\textwidth}
  \centering
    \includegraphics[width=1\linewidth,page=2]{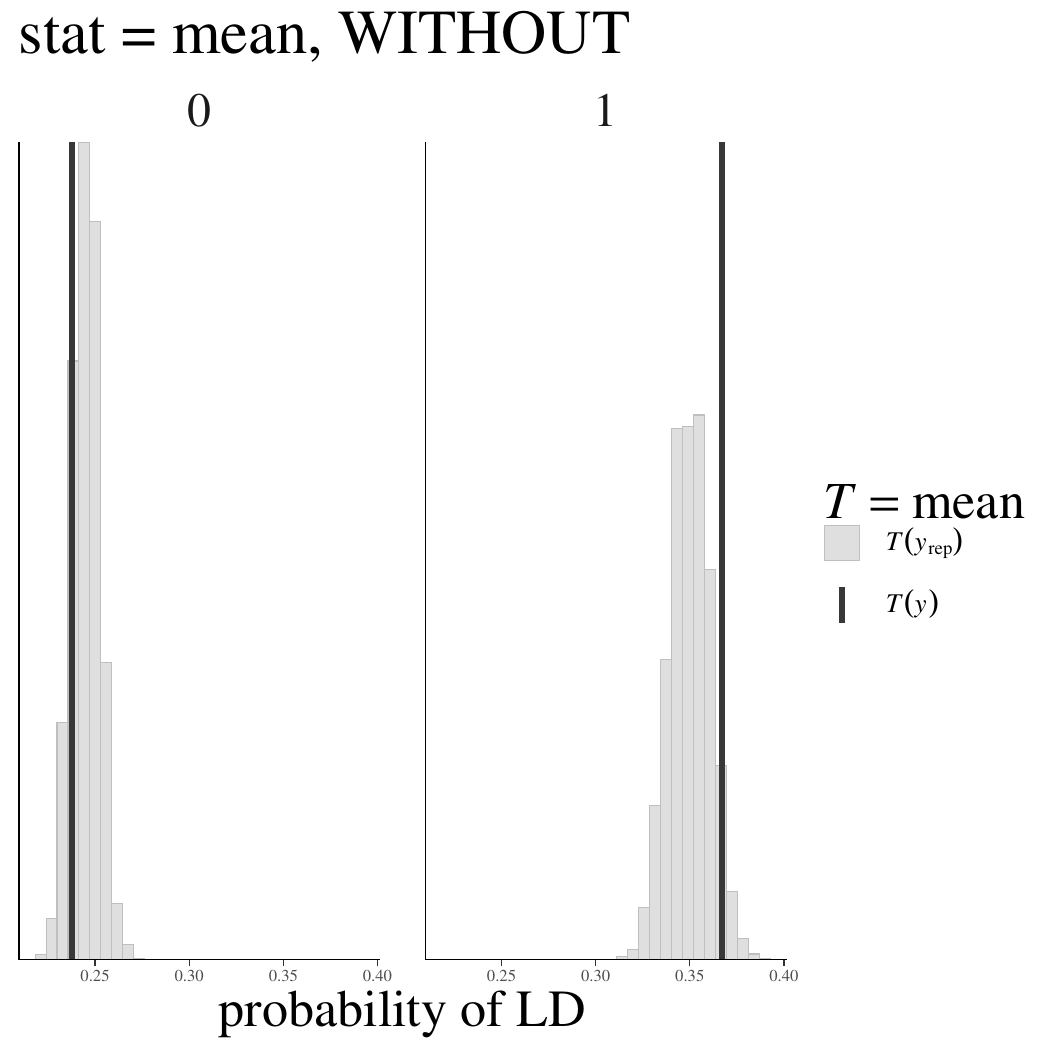}
    \caption{Candidate model 2}
  \end{subfigure}%
  \begin{subfigure}{.33\textwidth}
  \centering
    \includegraphics[width=1\linewidth,page=2]{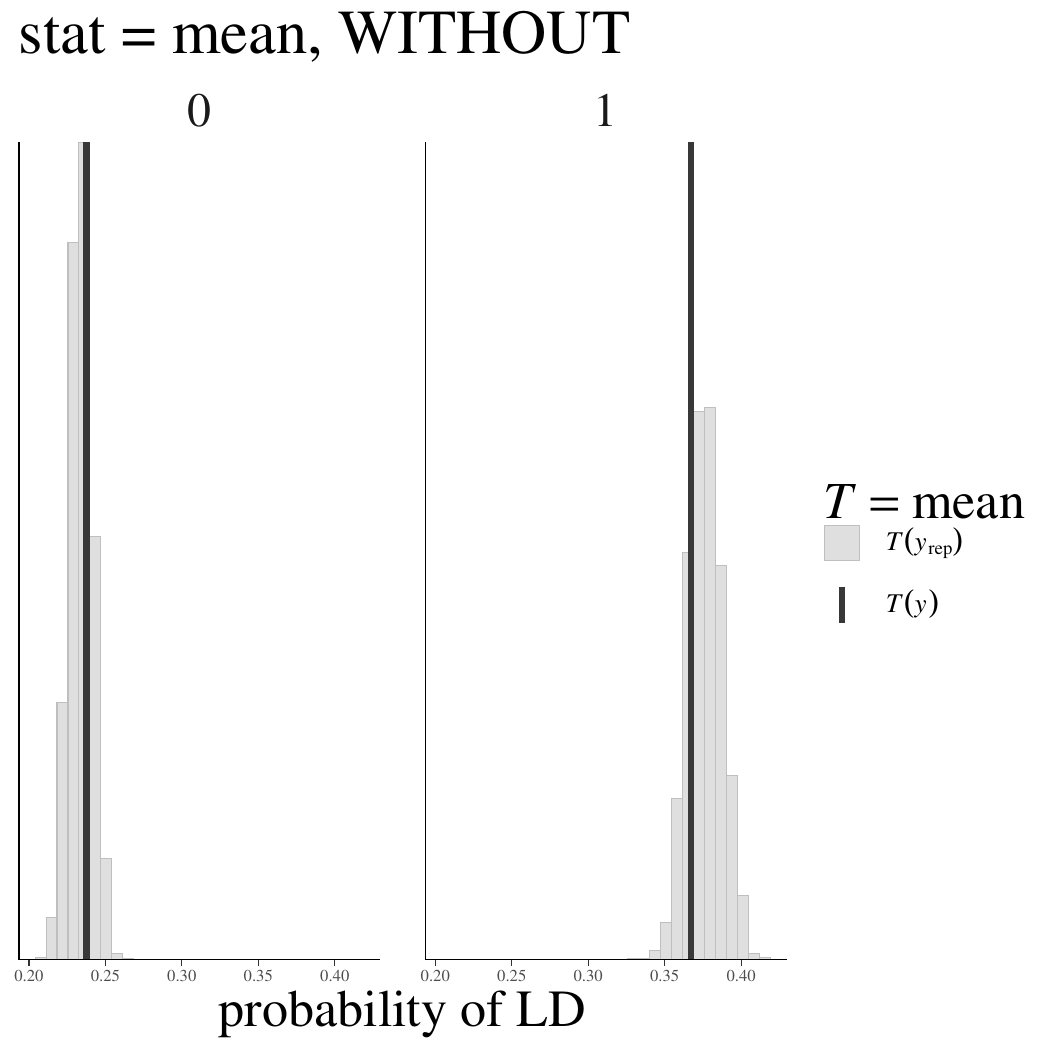}
    \caption{Candidate model 3}
  \end{subfigure}
  \begin{subfigure}{.33\textwidth}
  \centering
    \includegraphics[width=1\linewidth,page=2]{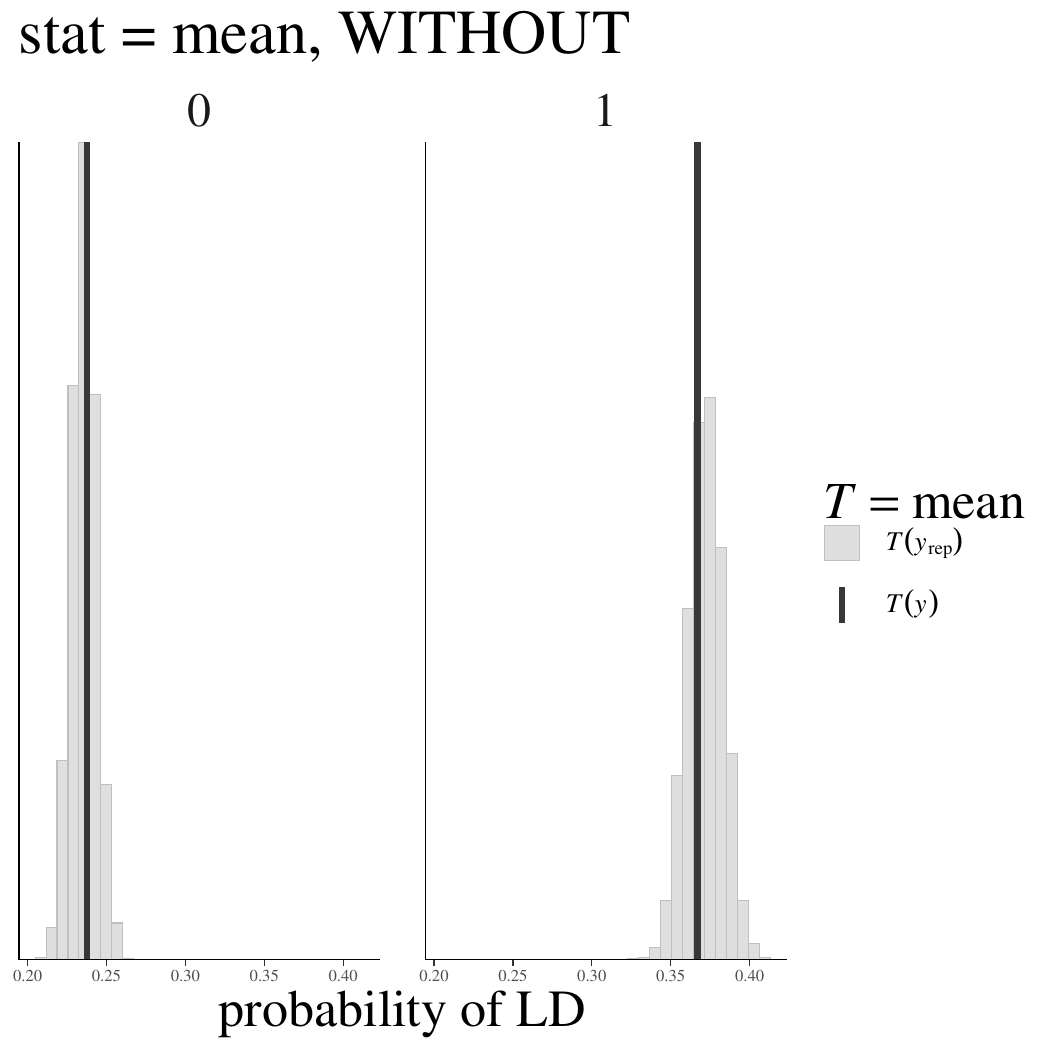}
    \caption{Candidate model 4}
  \end{subfigure}
  \caption{PPCs for the hold-out predictor \textit{WITHOUT},
  the left plot corresponds to \textit{WITHOUT = 0},
  the right plot to \textit{WITHOUT = 0}.\label{stack_ho1}}
\end{figure}

As already mentioned in section \ref{appppc}, we investigate the simulated posterior predictive distribution of the outcome variable for the binary predictor \textit{WITHOUT}, which was not used in either of the candidate models. If the PPCs show no anomaly for the subgroup means defined by \textit{WITHOUT}, we can assume that its explanatory value to literacy deprivation is captured by the predictors from the stacking ensemble. The variable \textit{WITHOUT} could be described as a "hold-out predictor", analogous to the terminology used in cross-validation for machine learning, when parts of the data are not used to "train" the model, and instead kept as "hold-out" for investigating the predictive performance of the model. Figure \ref{stack_ho1} displays the PPCs for literacy deprivation using the six models (stacking ensemble, core model and the four candidate models) conditioned on \textit{WITHOUT}. Again, the solid black line denotes the actually observed sample mean. The PPCs confirm the inadequacy of the core model, and demonstrate how the stacking ensemble passes this test, as the TSPPPV for the stacking ensemble is $0.68$ for \textit{WITHOUT = 0} and $0.63$ for \textit{WITHOUT = 1}. Among the candidate models, models 3 and 4 predict the corresponding region of the outcome distribution very well, too. In particular, candidate model 4 even outperforms the stacking ensemble and its posterior predictive distribution has a smaller variance. This reflects that the stack is more conservative and, thus, incorporates model uncertainty\footnote{The corresponding TSPPPV are given in tables \ref{tstack_ho1} in appendix \ref{app:step4}.}.
This suggests that the predictors used in models 3 and 4 are particularly important for predicting
the corresponding regions of the outcome distribution. Thus, the proportion of literacy deprived students among schools with either high or low dropout rates is accurately predicted by these models.

\par  In a real research process, the validity should be examined on the basis of several hold-out variables. The purpose here is to check whether the stacking ensemble provides valid predictions for different regions of the outcome distribution. The researcher has the flexibility to use several hold-out variables to examine validity accordingly.

\subsection{Step 5: Summarizing Results}

\label{step5}

\par The central component of any Bayesian analysis is the posterior distribution of the model parameters\footnote{Note that parameters are considered to be random variables in the Bayesian framework}, which also serves as the basis for all the findings we display in this subsection. A key benefit of Bayesian statistics is that the researcher can summarize the whole posterior distribution regarding the 
effect of substantive interest. This means in our analysis example that we summarize the posterior distribution of $p_{LD}$, the probability of being literacy deprived, for typical female and male students regarding their distribution of non-focal variables. We also derive the corresponding gender gap, i.e. the posterior distribution $f([p_{LD}|FEM=1]-[p_{LD}|FEM=0])$. Our summary of the substantively important results of the stacking ensembles is based on predictive comparisons, since this type of result presentation is ideal for summarizing the structure of a complex predictive model with regard to the research question \citep{GelmanPardoe}. Based on the posterior distribution of the stacking ensemble the researcher can derive conditional model-based predictions which are of substantive interest to them, and we used the complete list of predictors from the model specification to derive the conditional model-based predictions.

\par Foundation for all subsequent analyses are the posterior distributions of the probabilities of being literacy deprived conditioned on the eight possible combinations of the focal variables \textit{FEM} and \textit{SES}, where we set the remaining variables to specific values. The corresponding summaries and sample sizes can be found in table \ref{refgrid} of the appendix section \ref{app:step1}. These remaining variables are referred to  as "non-focal" variables in the following.

\par Prior to the model building process, we prepared the non-focal predictors such that it is plausible, regarding the literature, that high values are associated with a low risk of being literacy deprived.
Based on the $38$ non-focal variables we created three different hypothetical students by setting their respective values for each of the $38$ variables to the lower quartile, median and upper quartile of each non-focal variable. This allows us to account for the variation of the distribution of the non-focal variables within the eight specific subgroups of the focal variables. 

\par
\begin{figure}[H]
\centering
\includegraphics[width=7.5cm]{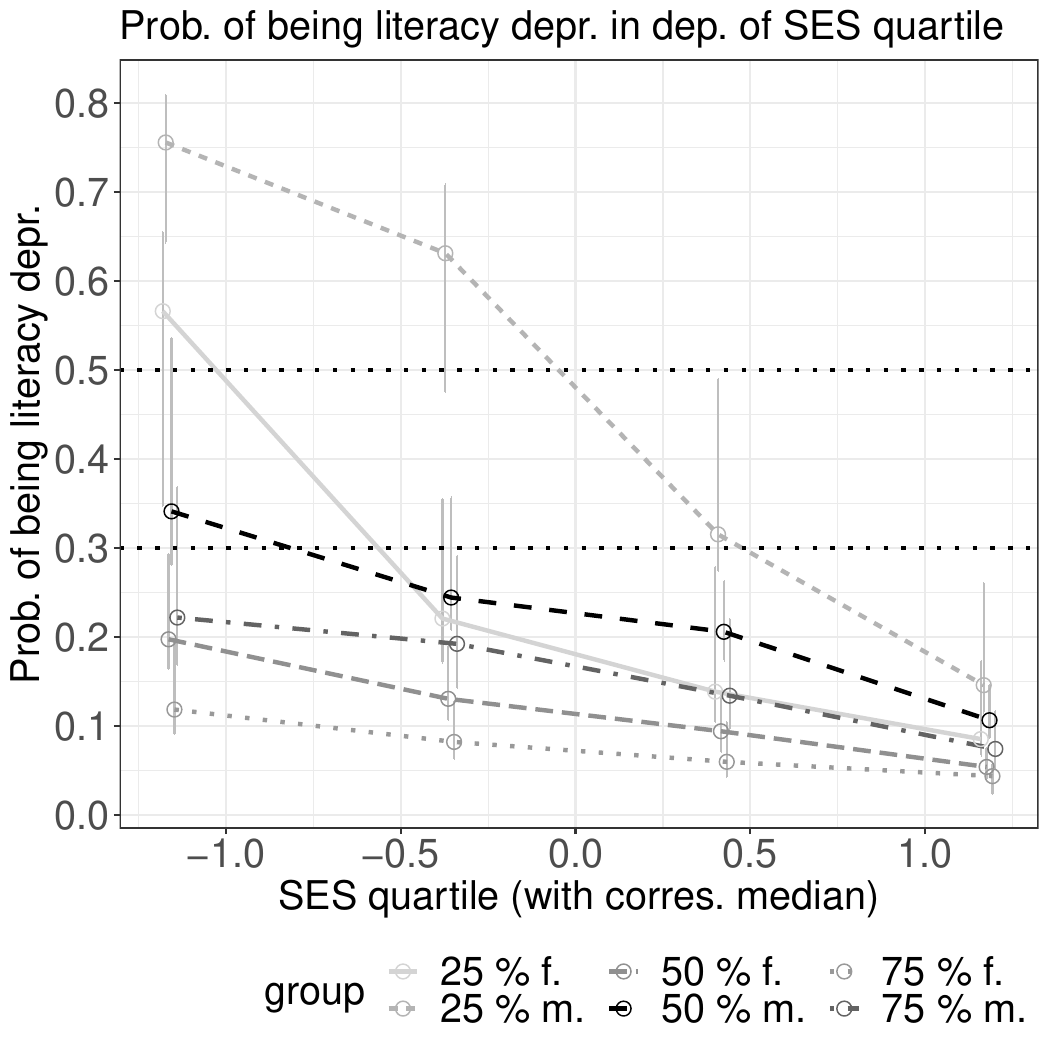}
\caption{\label{fig_apd1} 
\small{Predicted probability of being literacy deprived in dependence of the two focal predictors \textit{FEM} and \textit{SES} (\textit{f = female, m = male, $25\%$ corresponds to the hypothetical lower-quartile student, $50\%$ corresponds to the hypothetical median student, and $75\%$ corresponds to the hypothetical upper-quartile student of the non-focal variables})}}
\end{figure}

 Figure \ref{fig_apd1} displays an overview of the  posterior distributions of $p_{LD}$ for the $8\times 3=24$ hypothetical students, where the x-axis is the ESCS and the corresponding median for each SES quartile, and the y-axis is the predicted probability of being literacy deprived based on the stacking ensemble.\footnote{The corresponding table \ref{resultsPostDist} is given in the appendix \ref{app:step5}.}

The figure suggests that the specific values of the non-focal variables have, in general, considerable explanatory power regarding the risk of being literacy deprived, but this is particularly true for socio-economically disadvantaged students. Note that the differences displayed in figure \ref{fig_apd1} are not governed by some functional specification of the underlying model. In particular, the risk of being literacy deprived is quite high for the non-focal lower-quartile students in the bottom quartile of ESCS (for both, male and female students). The risk of being literacy deprived decreases considerably
with increasing ESCS quartiles for all hypothetical students, while at the same time variations in differences are decreasing as well. 
It is also evident that male students have a higher risk than the female comparison group, which is in line with the current state of research.

The primary research interest in this example is to measure the gender gap in literacy deprivation. We compute the corresponding gaps $[p_{LD}|FEM=1]-[p_{LD}|FEM=0]$ for the hypothetical median student (regarding the $38$ non-focal variables) as a show-case.\footnote{The other two cases, the lower and upper quartile of the non-focal variables, are given in  appendix \ref{app:step5}.} The corresponding results are displayed in table \ref{t4gap}. The first columns displays  the range of the corresponding ESCS quartile, and the second to fourth column the median, 5\% and 95\% quantile of the posterior distribution of the gender gap. The positive values across all quantiles indicate that the risk of being literacy deprived is in all four groups higher for male students. 
\begin{table}[!htb]
    \caption{Gender gaps for the group of students that corresponds to the median of the non-focal variables in figure \ref{fig_apd1}\label{t4gap}}
    \centering
      \begin{tabular}{rrrr}
 \hline

ESCS quartiles & post. median & post. 5\% quantile & post. 95\% quantile \\
\hline
[-6.68,-0.75]   & 0.15 & 0.08 & 0.30\\
(-0.75,0.0058]  & 0.12 & 0.08 & 0.18\\
(0.0058,0.822]  & 0.12 & 0.05 & 0.15\\
(0.822,3.57]   & 0.06 & 0.03 & 0.08\\
\hline
\end{tabular}
\end{table}

 Table \ref{t4gap} summarizes our posterior knowledge about the gender gap of being literacy deprived for the corresponding ESCS quartile, and it allow us to quantify the gender gap of being literacy deprived for that quartile. These gaps and corresponding intervals in table \ref{t4gap} reflect epistemic uncertainty \citep[see also][]{Kaplan_Harra_2023}. 

\par Our analysis demonstrated that with increasing level of socio-economic background the average literacy deprivation gender gap becomes noticeably smaller. For the group at the bottom quartile ESCS the posterior information about the gap includes a wide gender-specific risk difference. Taking into account the gender gaps for the lower and upper quartile of the non-focal variables, given in \ref{tgap25} and \ref{tgap75} in appendix \ref{app:step5}, we gain a deeper and more nuanced  understanding of the social phenomenon under investigation: The group of students at the upper quartile has a small gender gap
independent of the ESCS quartile. The biggest gender gap for this group is $10\%.$
For the hypothetical lower-quartile (of the non-focal predictors) student the gender gap is much wider, since it gets close to $45\%$.

\section{Extensions}
\label{sec:extensions}
This paper examined Bayesian stacking in the context of a relatively simple set of logistic regression models using data from the PISA 2018 study. If Bayesian stacking is to be used for the development of serious policy-relevant models (which are needed to guide national educational policy), new methodological developments would be required to account for the complexity of these assessments. We outline these extensions in this section.

In Section \ref{sec:EmpAppl} we briefly reviewed the structure of the PISA data set. It is clear that properly modeling data from PISA requires consideration of the multilevel nature of the data as well as the use of sampling weights to account for the complex nature of the design.  
It is beyond the scope and purpose of this study to address these issues, but each model in the stack could be specified as a Bayesian hierarchical model to account for PISA's multilevel structure. The basic idea is to assign hyperparameters and hyperprior distributions, which are functions of between-school predictors, to the parameters for the within-school model \citep{kaplanbayesbook2}. 
Stacking can then be easily extended to Bayesian hierarchical models, the researcher can just stack Bayesian hierarchical models. Moreover PPCs can be applied to Bayesian hierarchical  models as well and, thus, the workflow can be transferred without effort to deal with Bayesian hierarchical models. However, the calculation of the Bayesian hierarchical model and the calculation of the stacking weights takes both significantly longer. In addition, numerical problems can occur when calculating the distribution of the posterior parameters and the stacking weights. 

Sampling weights would require us to identify the appropriate weights, and software programs such as rstanarm and brms \citep{burkner_brms_2017} would need functionalities to incorporate all the weights into the model. Experience has shown that raw sampling weights can lead to unstable estimation of the likelihood and hence it may be useful to normalize the sampling weights ahead of their use \citep{Kaplan_Harra_2023}.

A recently proposed expansion to Bayesian stacking, namely hierarchical stacking \citep*{yao2021bayesian} is a more flexible approach than original stacking. A hyperprior is applied to the stacking weights and thus allows for more flexibility in the computation of the stacking ensemble which results in a more nuanced approach to capture the variability inherent in the data.
Therefore, the usage of hierarchical stacking in combination with multilevel models is a promising extension of our workflow.

\section{Summary and Discussion} \label{sec:Outlook}

\par Model uncertainty is pervasive in real world analysis situations and is still an often-neglected issue in applied statistics. There exists an infinite number of possibilities to make decisions in the model specification process, and each decision on specifying the prior and the likelihood affects the results. An overlooked aspect in model building that biases model selection towards simple models is the issue of interpretability, since there is a trade-off between flexibility and interpretability \citep*[see e.g.][25]{james_introduction_2021}. When models are used as a basis for policies (e.g. regarding the educational system), even slight differences in terms of model outcome can have a very high impact on real-world decisions. Therefore, it is sensible to deviate from the current practice to put all your eggs in one (model) basket, and instead explore the hypothetical answers which different, but almost equally plausible models, would give you and, thus, to reflect model uncertainty correctly. Based on the approaches outlined in this work and due to the availability of large data sets, researchers can gain insights when they interpret empirical patterns in the light of existing substantive knowledge or sociological theory, while taking into account model uncertainty. 

 \par Our work introduces a viable workflow for incorporating PPC into predictive regression modeling using Bayesian stacking. The analysis assumes a $\mathcal{M}-{open}$ framework and specifies a set of regression models with different predictors with the aim to predict distinct regions of the outcome distribution. These different regression models represent reasonable alternative candidate models without presuming the existence of a true model. Bayesian stacking then
provides a separate score for each candidate model. By using their scores as weights these models are combined to derive the combined predictive distribution. This predictive distribution is used to perform PPC on the one hand to assess the predictive quality of those elements of the stacking ensemble that are important to the research question and on the other hand to investigate and understand the validity of the stacking ensemble to make predictions for relevant groups of students.

\par Predictive modeling can advance explanatory ambitions in the social sciences \citep{GalitShmueli.2010} and in addition enrich the understanding of social phenomena. Our workflow provides a show-case example of how Bayesian stacking and PPC can be used for this purpose. We demonstrate the applicability of our workflow by conducting an analysis for the effect of the gender gap on being literacy deprived using large-scale assessment data from the PISA study. The approach is in particular promising for large-scale assessment trend analysis, because it allows for investigation of conditional factors among known drivers of social inequality. One key advantage is the flexibility of the approach: The nature of the complex pattern displayed in figure \ref{fig_apd1} does not rely on a model assumption that typically implies some monotone form of interaction, but can instead be investigated comparing the predictive distributions for different factor combinations, which might yield non-monotone associations.
 
 \par Moreover, our proposed workflow relies on predictive modeling rather than direct parameter interpretation and therefore circumvents the tendency towards simpler models. In combination with our suggestions regarding model uncertainty we believe that we have provided a useful toolbox to make statistical inference more stable and less arbitrary. While it is also technically possible to incorporate model uncertainty in frequentist statistics, the integration is far more intuitive within the Bayesian framework, since all that is required is the inclusion of the model as another condition. Ironically, Bayesian inference is sometimes criticized for being subjective, as it requires the specification of a prior distribution that reflects initial subjective belief about the model parameters, but the subjectivity that is connected to the process of selecting a single "final" model has never had to face comparative levels of criticism and has been largely swept under the rug. Our hope is that the present work, and those of others cited in this paper push the problem of model uncertainty more to the forefront of statistical modeling in the social sciences.

\newpage
\appendix

\newpage

\clearpage

\section{Univariate Summary Statistics for Focal Predictors (Step 1)}
\label{univ}
\begin{table}[h!]
    \caption{\small{Quartiles of the standardized ESCS, the first line defines the 
    corresponding ESCS range, the second line indicates the number of the \textit{SES} and the third line informs about the size of respective subpopulation,
    the forth line is the median ESCS of the corresponding subpopulation.} \vspace{0.2cm} }
    \centering
\begin{tabular}{|r|r|r|r|r|}
\hline
\textit{ESCS quartiles}  &  [-6.68,-0.75]  & (-0.75,0.0058] & (0.0058,0.822]&  (0.822,3.57] \\
\hline
\textit{SES}  &  1 & 2 &  3 &  4 \\ \hline
sample size & 1509 & 1555 & 1580 &   1584\\
\hline
median ESCS & -1.163  &  -0.361 &  0.422  &  1.182 \\ %    
\hline
\end{tabular}
\label{tab:ESCS4}
\end{table}

\label{app:step1}

\begin{table}[H]
\begin{center}

    \caption{\small{This table (the focal grid) provides an overview by forming a grid from the two focal variables and indicating the proportion of literary deprived students fr(LD) and the group size in the respective case. \label{refgrid}}\vspace{0.2cm}}
    \centering

\begin{tabular}{|r|r|r|r|}
\hline
FEM & SES & fr(LD) & Size\\
\hline
0 & 1 & 0.518 & 736\\
\hline
0 & 2 & 0.402 & 792\\
\hline
0 & 3 & 0.332 & 793\\
\hline
0 & 4 & 0.174 & 793\\
\hline
1 & 1 & 0.321 & 773\\
\hline
1 & 2 & 0.224 & 763\\
\hline
1 & 3 & 0.159 & 787\\
\hline
1 & 4 & 0.096 & 791\\
\hline
\end{tabular}

\label{tab:focGrid}

\end{center}
\end{table}

\section{Recoding and Imputation}

Several variables were transformed or recoded when the data were prepared for the modeling steps described in section \ref{sec:Workflow}. Since these preparations are very data specific and do not translate from one analysis to another, they were omitted from the paper.

Some parts of the data were also suffering from nonresponse and we decided to use the R package \emph{mice}\citep{vanBuuren.2011} to impute the missing values. The algorithm is based on sequential regressions and therefore suited to handle the non-monotone missing-data pattern. We used the mice imputation method 'cart', which is based on classification and regression trees, and considered to be robust to model misspecification. The sequential regression step was repeated only twice due to the sheer size of the data set.

Although mice is a well-established multiple imputation\citep{Rubin.1987} method, we decided to use single imputation, and treat the data for our demonstration of the workflow as if the data had been complete. While multiple imputation accounts, which accounts for uncertainty due to missing information, is using the Bayesian paradigm itself in the imputation phase, the subsequent analysis is frequentist. Doing Bayesian analysis on multiply imputed data has been proposed \citep[see][]{zhou_note_2010}, the procedure is not straightforward and would have distracted from our actual research goal. The same is true for incorporating the imputation step via data augmentation within the Hamiltonian Monte Carlo phase, since the stan-based convenience functions from the \emph{rstanarm} package to not include this option, and we would have had to program this step in stan.

\section{Variable Definition for Core and Candidate Models (Step 2)}
\label{app:step2}
Some scales are reversed, so that high values plausibly are associated with a low risk of literacy deprivation. The scales are indicated with (r).
\begin{table}[H]

\caption{Overview of the core  model. \vspace{0.2 cm}}
\centering
\begin{tabular}[t]{|c|l|l|}
\hline
Predictor & Description & PISA code\\
\hline
(Intercept) & Intercept & / \\
\hline
FEM & Gender (binary, 1 = reported female) & ST004 \\
\hline
SES2 & Second quartile of ESCS & ESCS\\
\hline
SES3 & Third quartile of ESCS &  ESCS \\
\hline
SES4 & Top quartile of ESCS &  ESCS\\
\hline
\end{tabular}
\label{tab:Varoverview_c0}

\end{table}

\begin{table}[H]

    \caption{\small{Overview of the first candidate model, core model predictors plus the following predictors: } }
    \centering
    \begin{tabular}{|c|l|l|}
     \hline
    Predictor & Description & PISA Code\\
        \hline
        \
COMP & Perception of reading competence  & ST161\\
\hline
NDIFF & Perception of reading difficulty (r) & ST161\\
\hline
JOY & Enjoyment of reading & ST160\\
\hline
JOYP\_SC & School mean parent enjoyment of reading & PA158\\
\hline
COMP:NDIFF & Interaction & /\\
\hline
FEM:JOY & Interaction & /\\
\hline
SES2:NDIFF & Interaction & /\\
\hline
SES3:NDIFF & Interaction & /\\
\hline
SES4:NDIFF & Interaction & /\\  
\hline
FEM:SES2 & Interaction & /\\
\hline
FEM:SES3 & Interaction & /\\
\hline
FEM:SES4 & Interaction & /\\ 
\hline
\end{tabular}
\label{tab:Varoverview_c1}

\end{table}

\begin{table}[h!]

    \caption{\small{Overview of the second candidate model, core model predictors plus the following predictors:} }
    \centering
    \begin{tabular}{|c|l|l|}
    \hline
    Variable label & Description & PISA Code\\
 \hline
 UNI &  Expected graduation, (binary, 1 if ISCED level 5 A or 6) & ST225\\
\hline
VALS & Value of school & ST036\\
\hline
RES & Self-efficacy & ST188\\
\hline
TF & Teacher feedback & ST104\\
\hline
TSUPP & Teacher support & ST123\\
\hline
TDIN & Teacher-directed instruction & ST102\\
\hline
ATT & Attendance problem (regular lateness or absence) & ST062 \\
\hline
GFOF & Fear of Failure & ST183\\
\hline
MOT & Motivation to master tasks & ST182\\ 
\hline
NBULLV &  Bullying victim (binary, 1 = no)  & ST038\\
\hline
MOT\_SC & Mean school motivation to master task & ST182\\
\hline
GOAL\_SC & School mean learning goals & ST208\\
\hline
SES2:MOT\_SC & Interaction & /\\
\hline
SES3:MOT\_SC & Interaction & /\\
\hline
SES4:MOT\_SC & Interaction & /\\ 
\hline
FEM:UNI& Interaction & /\\
\hline
FEM:SES2 & Interaction & /\\
\hline
FEM:SES3 & Interaction & /\\
\hline
FEM:SES4 & Interaction & /\\ 
\hline
\end{tabular}
\label{tab:Varoverview_c3}

\end{table}

\clearpage

\begin{table}[H]

    \caption{\small{Overview of the third candidate model, core model predictors plus the following predictors:} }
    \centering
    \begin{tabular}{|c|l|l|}
     \hline
    Variable label & Description & PISA Code\\
 \hline
    ACA & Academic household (binary, 1 = yes)  & HISCED\\
\hline
NATIVE & Language at home is test language (binary, 1 = yes) & ST022\\
\hline
NIMMIG & Immigration background  (binary, 1 = no)  & IMMIG \\
\hline
NFEW\_BOOKS & 26 books maximum (binary, 1 = no) & ST013 \\ % minimum
\hline
NSN\_SC & Ab. med. perc. of stud. with spec. needs (binary, 1 = no) & SC048 \\
\hline
N\_SC & School fraction of natives  & SC048 \\
\hline
NDISHOME & Disadvantaged area school (binary, 1 = no) & SC048Q03NA \\
\hline
UNI\_SC & School mean of UNI & ST225 \\
\hline
NATIVE:NIMMIG & Interaction & / \\
\hline
SES2:ACA & Interaction & / \\
\hline
SES3:ACA & Interaction & /\\
\hline
SES4:ACA & Interaction & /\\
\hline
FEM:UNI\_SC & Interaction & / \\
\hline
ACA:NDISHOME & Interaction & /\\
\hline
\end{tabular}
\label{tab:Varoverview_c2}

\end{table}

\begin{table}[h!]

    \caption{\small{Overview of the forth candidate model, core model predictors plus the following predictors:} }
    \centering
    \begin{tabular}{|c|l|l|}
    \hline
    Variable label & Description & PISA code\\\hline
    
BEL & Sense of Belonging to school (binary, 1 = yes) & ST034\\
\hline
STAFF &  Shortage of staff above median (binary, 1 = no)& SC017\\
\hline
EDU &   Shortage of educ. material above median (binary, 1 = no) & SC017\\
\hline
NFEWB\_SC & Sc. prop. of students with no book shortage & ST013\\
\hline
ACA\_SC & Sc. prop. of students from acad. housh. & HISCED \\
\hline
NBULL\_SC & Sc. prop. of students without sev. bull. exp. & ST038\\
\hline
TDIN\_SC & Sc. mean of teacher-directed instruction & ST102 \\
\hline
TS\_SC  & Sc. mean of teacher support & ST123 \\
\hline
TF\_SC &  Sc. mean of teacher feedback & ST104 \\
\hline
RES\_SC & Sc. mean of student self-efficacy & ST188 \\
\hline
STUBEHA & Index of students behaviour hindering learning & SC061 \\
\hline
TEACHBEHA & Index of teachers behaviour hindering learning & SC061 \\
\hline
DC\_SC & Sc. mean disciplinary climate & ST097  \\
\hline
NRUR & Rural School (binary, 1 = no) & SC0\\
\hline
NFEWB\_SC:NRUR & Interaction & / \\
\hline
FEM:NRUR & Interaction & /  \\
\hline
SES2:NBULL\_SC & Interaction & / \\
\hline
SES3:NBULL\_SC & Interaction & / \\
\hline
SES4:NBULL\_SC &Interaction & /  \\
\hline

\end{tabular}
\label{tab:Varoverview_c4}
\end{table}

\section{Regression Results (Step 3) }
\label{regRes}

\begin{table}
\caption{Regression coefficients of the core model}
\centering
\begin{tabular}[t]{|l|r|r|r|r|}
\hline
term & estimate & std.error & hpd.low & hpd.high\\
\hline
(Intercept) & 0.0820 & 0.0606 & -0.0352 & 0.2021\\
\hline
FEM & -0.8417 & 0.0585 & -0.9669 & -0.7218\\
\hline
SES2 & -0.4808 & 0.0805 & -0.6350 & -0.3305\\
\hline
SES3 & -0.8310 & 0.0793 & -0.9851 & -0.6693\\
\hline
SES4 & -1.5818 & 0.0907 & -1.7649 & -1.4131\\
\hline
\end{tabular}
\end{table}

\clearpage

\begin{table}[H]
\caption{Regression coefficients of the candidate model  1}
\centering
\begin{tabular}[t]{|l|r|r|r|r|}
\hline
term & estimate & std.error & hpd.low & hpd.high\\
\hline
(Intercept) & -0.2483 & 0.0832 & -0.4142 & -0.0826\\
\hline
FEM & -0.6455 & 0.1151 & -0.8750 & -0.4137\\
\hline
SES2 & -0.3413 & 0.1127 & -0.5585 & -0.1192\\
\hline
SES3 & -0.6436 & 0.1121 & -0.8764 & -0.4113\\
\hline
SES4 & -1.2811 & 0.1259 & -1.5342 & -1.0370\\
\hline
COMP & -0.3172 & 0.0358 & -0.3849 & -0.2514\\
\hline
NDIFF & -0.3515 & 0.0641 & -0.4752 & -0.2304\\
\hline
JOY & 0.0237 & 0.0486 & -0.0685 & 0.1200\\
\hline
JOYP\_SC & -1.6536 & 0.1110 & -1.8854 & -1.4249\\
\hline
COMP:NDIFF & -0.1345 & 0.0253 & -0.1846 & -0.0833\\
\hline
FEM:JOY & -0.3883 & 0.0718 & -0.5333 & -0.2485\\
\hline
SES2:NDIFF & -0.1076 & 0.0918 & -0.2798 & 0.0693\\
\hline
SES3:NDIFF & -0.1547 & 0.0913 & -0.3324 & 0.0196\\
\hline
SES4:NDIFF & -0.0879 & 0.1025 & -0.2879 & 0.1102\\
\hline
FEM:SES2 & -0.0776 & 0.1647 & -0.3988 & 0.2508\\
\hline
FEM:SES3 & -0.0368 & 0.1750 & -0.3864 & 0.3016\\
\hline
FEM:SES4 & 0.1828 & 0.1992 & -0.1969 & 0.5693\\
\hline
\end{tabular}
\end{table}

\begin{table}[H]
\caption{Regression coefficients of the candidate model  2 }
\centering
\begin{scriptsize}
\begin{tabular}[t]{|l|r|r|r|r|}
\hline
term & estimate & std.error & hpd.low & hpd.high\\
\hline
(Intercept) & 0.3198 & 0.1057 & 0.1169 & 0.5267\\
\hline
FEM & -0.2732 & 0.1546 & -0.5754 & 0.0240\\
\hline
SES2 & -0.2780 & 0.1238 & -0.5189 & -0.0313\\
\hline
SES3 & -0.3725 & 0.1273 & -0.6106 & -0.1302\\
\hline
SES4 & -0.9554 & 0.1369 & -1.2387 & -0.6946\\
\hline
UNI & -1.2704 & 0.0953 & -1.4579 & -1.0782\\
\hline
VALS & 0.0743 & 0.0342 & 0.0041 & 0.1440\\
\hline
RES & -0.1159 & 0.0385 & -0.1887 & -0.0410\\
\hline
TF & 0.0767 & 0.0362 & 0.0076 & 0.1468\\
\hline
TSUPP & -0.1702 & 0.0438 & -0.2528 & -0.0873\\
\hline
TDIN & 0.2023 & 0.0423 & 0.1191 & 0.2872\\
\hline
ATT & -0.3285 & 0.0322 & -0.3901 & -0.2639\\
\hline
GFOF & -0.0066 & 0.0344 & -0.0772 & 0.0619\\
\hline
MOT & -0.1797 & 0.0386 & -0.2532 & -0.1054\\
\hline
NBULLV & 0.7166 & 0.0934 & 0.5243 & 0.9000\\
\hline
MOT\_SC & -1.4042 & 0.2665 & -1.9313 & -0.8823\\
\hline
GOAL\_SC & -0.4941 & 0.1748 & -0.8323 & -0.1597\\
\hline
SES2:MOT\_SC & 0.3878 & 0.3569 & -0.3204 & 1.0678\\
\hline
SES3:MOT\_SC & 0.0697 & 0.3708 & -0.6382 & 0.7936\\
\hline
SES4:MOT\_SC & 0.1750 & 0.4146 & -0.6073 & 0.9938\\
\hline
FEM:UNI & -0.2987 & 0.1495 & -0.5927 & -0.0039\\
\hline
FEM:SES2 & -0.0597 & 0.1803 & -0.4154 & 0.2859\\
\hline
FEM:SES3 & -0.2237 & 0.1850 & -0.5916 & 0.1350\\
\hline
FEM:SES4 & 0.1498 & 0.2072 & -0.2588 & 0.5680\\
\hline
\end{tabular}
\end{scriptsize}
\end{table}

\begin{table}[H]
\caption{Regression coefficients of the candidate model  3}
\centering
\begin{tabular}[t]{|l|r|r|r|r|}
\hline
term & estimate & std.error & hpd.low & hpd.high\\
\hline
(Intercept) & 3.7221 & 0.2611 & 3.2164 & 4.2663\\
\hline
FEM & -0.3581 & 0.2732 & -0.9035 & 0.1601\\
\hline
SES2 & -0.2610 & 0.0953 & -0.4522 & -0.0703\\
\hline
SES3 & -0.1006 & 0.1297 & -0.3437 & 0.1449\\
\hline
SES4 & 0.5205 & 0.2922 & -0.0705 & 1.1009\\
\hline
ACA & 0.7564 & 0.3360 & 0.1328 & 1.3920\\
\hline
NATIVE & -0.4395 & 0.1850 & -0.7934 & -0.0774\\
\hline
NIMMIG & 0.3907 & 0.3231 & -0.2399 & 0.9947\\
\hline
NFEW\_BOOKS & -0.5661 & 0.0743 & -0.7108 & -0.4126\\
\hline
NSN\_SC & -0.0858 & 0.0671 & -0.2181 & 0.0492\\
\hline
N\_SC & -0.0003 & 0.1701 & -0.3386 & 0.3359\\
\hline
NDISHOME & 0.0742 & 0.0903 & -0.0991 & 0.2454\\
\hline
UNI\_SC & -4.3027 & 0.2324 & -4.7644 & -3.8521\\
\hline
NATIVE:NIMMIG & -0.5976 & 0.3542 & -1.2391 & 0.0736\\
\hline
SES2:ACA & -0.1755 & 0.3472 & -0.8907 & 0.4925\\
\hline
SES3:ACA & -0.8846 & 0.3526 & -1.5791 & -0.1997\\
\hline
SES4:ACA & -2.0010 & 0.4343 & -2.8691 & -1.1365\\
\hline
FEM:UNI\_SC & -0.4852 & 0.3576 & -1.1913 & 0.2138\\
\hline
ACA:NDISHOME & -0.0420 & 0.1347 & -0.3073 & 0.2189\\
\hline
\end{tabular}
\end{table}

\begin{table}[H]
\caption{Regression coefficients of the candidate model  4}
\centering
\begin{scriptsize}
\begin{tabular}[t]{|l|r|r|r|r|}
\hline
term & estimate & std.error & hpd.low & hpd.high\\
\hline
(Intercept) & 1.0291 & 0.3136 & 0.4014 & 1.6526\\
\hline
FEM & -0.9349 & 0.1132 & -1.1523 & -0.7114\\
\hline
SES2 & -0.4087 & 0.1511 & -0.7165 & -0.1093\\
\hline
SES3 & -0.9002 & 0.1720 & -1.2193 & -0.5778\\
\hline
SES4 & -1.3836 & 0.1891 & -1.7554 & -1.0193\\
\hline
BEL & -0.4317 & 0.0661 & -0.5620 & -0.3025\\
\hline
STAFF & -0.0016 & 0.0826 & -0.1663 & 0.1714\\
\hline
EDU & 0.1756 & 0.0696 & 0.0394 & 0.3050\\
\hline
NFEWB\_SC & -1.8860 & 0.4552 & -2.7404 & -0.9872\\
\hline
ACA\_SC & 0.2752 & 0.2943 & -0.2999 & 0.8624\\
\hline
NBULL\_SC & 2.0248 & 0.6256 & 0.7707 & 3.2618\\
\hline
TDIN\_SC & 0.7849 & 0.1943 & 0.4009 & 1.1622\\
\hline
TS\_SC & -0.4782 & 0.2063 & -0.8834 & -0.0815\\
\hline
TF\_SC & 0.2488 & 0.1679 & -0.0784 & 0.5620\\
\hline
RES\_SC & -0.3071 & 0.1877 & -0.6909 & 0.0581\\
\hline
STUBEHA & -0.0371 & 0.0439 & -0.1238 & 0.0546\\
\hline
TEACHBEHA & -0.0496 & 0.0445 & -0.1366 & 0.0403\\
\hline
DC\_SC & -0.8228 & 0.1163 & -1.0558 & -0.5968\\
\hline
NRUR & 0.9394 & 0.3588 & 0.2664 & 1.6430\\
\hline
NFEWB\_SC:NRUR & -1.7629 & 0.5004 & -2.7204 & -0.8304\\
\hline
FEM:NRUR & 0.2423 & 0.1405 & -0.0383 & 0.5100\\
\hline
SES2:NBULL\_SC & 1.1687 & 0.8533 & -0.4685 & 2.8641\\
\hline
SES3:NBULL\_SC & 3.7234 & 0.9461 & 1.9248 & 5.5097\\
\hline
SES4:NBULL\_SC & 4.0056 & 1.1539 & 1.8355 & 6.1912\\
\hline
\end{tabular}
\end{scriptsize}
\end{table}
\clearpage

\section{Posterior Predictive Checks (Step 4)}
\label{app:step4}
Two sided posterior predictive $p$-value for hold-out predictor
\textit{WITHOUT}

\begin{table}[!htb]
    \caption{Two sided posterior predictive $p$-values (TSPPPVs) that correspond to the graphical PPCs in figure \ref{stack_ho1}, above\textit{WITHOUT = 0}, below \textit{WITHOUT = 1}.\label{tstack_ho1}}
    \begin{subtable}{.33\linewidth}
      \centering
        \caption{Stacking ensemble}
 \begin{tabular}{|r|r|}
\hline
\textit{WITHOUT = 0} & 0.68\\
\hline
\textit{WITHOUT = 1} & 0.63\\
\hline
\end{tabular}
    \end{subtable}
    \begin{subtable}{.33\linewidth}
      \centering
        \caption{ Core model}
\begin{tabular}{|r|r|}
\hline
\textit{WITHOUT = 0} & 0.00\\
\hline
\textit{WITHOUT = 1} & 0.00\\
\hline
\end{tabular}
    \end{subtable} 
        \begin{subtable}{.33\linewidth}
      \centering
        \caption{Candidate model 1}
\begin{tabular}{|r|r|}
\hline
\textit{WITHOUT = 0} & 0.19\\
\hline
\textit{WITHOUT = 1} & 0.03\\
\hline
\end{tabular}
    \end{subtable} 

    \begin{subtable}{.33\linewidth}
      \centering
        \caption{Candidate model 2}
 \begin{tabular}{|r|r|}
\hline
\textit{WITHOUT = 0} & 0.32\\
\hline
\textit{WITHOUT = 0} & 0.10\\
\hline
\end{tabular}
    \end{subtable}
    \begin{subtable}{.33\linewidth}
      \centering
        \caption{Candidate model 3}
\begin{tabular}{|r|r|}
\hline
\textit{WITHOUT = 0} & 0.63\\
\hline
\textit{WITHOUT = 0} & 0.43\\
\hline
\end{tabular}
    \end{subtable} 
        \begin{subtable}{.33\linewidth}
      \centering
        \caption{Candidate model 4}
\begin{tabular}{|r|r|}
\hline
\textit{WITHOUT = 0} & 0.81\\
\hline
\textit{WITHOUT = 0} & 0.70\\
\hline
\end{tabular}
    \end{subtable} 
\end{table}

\section{Results (Step 5)}
\label{app:step5}

\begin{table}[H]

\caption{\footnotesize{Numerical summary of the post. dist. of the $24$ repr. stud. shown in \ref{fig_apd1}. The first columns inform about the median value of the ESCS for the corresponding ESCS quartile, the second column shows the median, the third column the 5\% quantile and the last column  the 95\% quantile of the post. dist. \label{resultsPostDist}}}
\begin{scriptsize}
\centering
\begin{tabular}[t]{|l|r|r|r|r|}
\hline
median(ESCS) & group & post. median & post. 5\% quant. & post. 95\% quant.\\ \hline
-1.16 & 25 \% m. & 0.769 & 0.643 & 0.813\\
\hline
-0.36 & 25 \% m. & 0.654 & 0.475 & 0.729\\
\hline
0.42 & 25 \% m. & 0.331 & 0.284 & 0.489\\
\hline
1.18 & 25 \% m. & 0.155 & 0.126 & 0.260\\
\hline
-1.16 & 25 \% f. & 0.590 & 0.347 & 0.681\\
\hline
-0.36 & 25 \% f. & 0.229 & 0.182 & 0.355\\
\hline
0.42 & 25 \% f. & 0.144 & 0.112 & 0.278\\
\hline
1.18 & 25 \% f. & 0.088 & 0.068 & 0.173\\
\hline
-1.16 & 50 \% m. & 0.337 & 0.282 & 0.536\\
\hline
-0.36 & 50 \% m. & 0.244 & 0.208 & 0.356\\
\hline
0.42 & 50 \% m. & 0.206 & 0.173 & 0.263\\
\hline
1.18 & 50 \% m. & 0.106 & 0.087 & 0.146\\
\hline
-1.16 & 50 \% f. & 0.196 & 0.164 & 0.293\\
\hline
-0.36 & 50 \% f. & 0.131 & 0.106 & 0.190\\
\hline
0.42 & 50 \% f. & 0.094 & 0.070 & 0.137\\
\hline
1.18 & 50 \% f. & 0.054 & 0.040 & 0.075\\
\hline
-1.16 & 75 \% m. & 0.212 & 0.160 & 0.368\\
\hline
-0.36 & 75 \% m. & 0.180 & 0.129 & 0.291\\
\hline
0.42 & 75 \% m. & 0.124 & 0.088 & 0.220\\
\hline
1.18 & 75 \% m. & 0.070 & 0.048 & 0.117\\
\hline
-1.16 & 75 \% f. & 0.112 & 0.082 & 0.198\\
\hline
-0.36 & 75 \% f. & 0.079 & 0.059 & 0.128\\
\hline
0.42 & 75 \% f. & 0.056 & 0.041 & 0.104\\
\hline
1.18 & 75 \% f. & 0.042 & 0.024 & 0.057\\
\hline
\end{tabular}
\end{scriptsize}
\end{table}

\begin{table}[!htb]
    \caption{\footnotesize{Gender gaps for the group of students that corresponds to the
$25\%$ quantile of the non-focal variables in figure \ref{fig_apd1}. The first column inform about the range of the ESCS for the corresponding ESCS quartile, the second column shows the median of the posterior distribution of the gender gap, the third column informs about the 5\% quantile of the posterior distribution and the last column informs about the 95\% quantile.}\label{tgap25}}
    \centering
      \begin{tabular}{rrrr}
 \hline
ESCS quartiles & post. median & post. 5\% quantile & post. 95\% quantile \\
\hline
[-6.68,-0.75]   & 0.19 & 0.10 & 0.35\\

(-0.75,0.0058] & 0.43 & 0.21 & 0.51\\

(0.0058,0.822] & 0.18 & 0.13 & 0.23\\

(0.822,3.57]  & 0.06 & 0.03 & 0.10\\
\hline
\end{tabular}
\end{table}

\begin{table}[!htb]
    \caption{\footnotesize{Gender gaps for the group of students that corresponds to the
$75\%$ quantile of the non-focal variables in figure \ref{fig_apd1}. The first column inform about the range of the ESCS for the corresponding ESCS quartile, the second column shows the median of the posterior distribution of the gender gap, the third column informs about the 5\% quantile of the posterior distribution and the last column informs about the 95\% quantile.}\label{tgap75}}
    \centering
      \begin{tabular}{rrrr}
 \hline
ESCS quartiles & post. median & post. 5\% quantile & post. 95\% quantile \\ \hline

[-6.68,-0.75]   & 0.11 & 0.06 & 0.19\\

(-0.75,0.0058]  & 0.11 & 0.07 & 0.20\\

(0.0058,0.822]  & 0.07 & 0.04 & 0.13\\

(0.822,3.57]   & 0.03 & 0.01 & 0.06\\
\hline
\end{tabular}
\end{table}

\bibliographystyle{agsm}

\renewcommand{\baselinestretch}{1.0}
\newpage
\clearpage

\bibliography{Pisa3}
\end{document}